\documentclass[a4paper,11pt]{article}
\pdfoutput=1
\usepackage{jcappub}

\compress

\usepackage{amsmath,amsfonts,amssymb,mathrsfs,graphicx,color,longtable,bm}
\usepackage{hyperref}
\usepackage{graphicx}
\usepackage{array}
\usepackage{cases} 
\usepackage[normalem]{ulem}
\usepackage[explicit]{titlesec}
\usepackage[utf8]{inputenc}
\usepackage[normalem]{ulem}
\usepackage{xcolor}
\usepackage{multirow}

\usepackage{placeins}

\usepackage{cleveref}
\crefname{figure}{figure}{figures}
\Crefname{figure}{Figure}{Figures}

\hypersetup{
     colorlinks   = true,
     citecolor    = red,
     urlcolor     = blue,
     linkcolor    = blue
}

\newcommand{\dd}{\mathrm{d}}

\makeatletter
\gdef\@fpheader{}
\makeatother
\usepackage{physics}
\usepackage{lipsum}
\usepackage{subfigure}
\usepackage{verbatim}

\newcommand{\noleft}{\left.\kern-\nulldelimiterspace}
\newcommand{\noright}{\right.\kern-\nulldelimiterspace}
\newcommand{\round}[1]{\left( #1 \right)}
\newcommand{\sround}[1]{\left[ #1 \right]}

\title{Signals from the early Universe: primordial features from the full-shape analysis of BOSS galaxies and eBOSS quasars in combination with Planck}

\author[a,b,c,d]{Mario Ballardini,}
\author[e]{Nicola Barbieri,}
\author[a,b]{Antonio Raffaelli,}
\author[f,g,h]{Chiara Moretti,}
\author[i,j]{Oliver H.~E.~Philcox,}
\author[f,g,h]{Emiliano Sefusatti,}
\author[k,l,f,g,h]{Francesco Verdiani}

\affiliation[a]{Dipartimento di Fisica e Scienze della Terra, Universit\`a degli Studi di Ferrara, via Giuseppe Saragat 1, 44122 Ferrara, Italy}
\affiliation[b]{INFN, Sezione di Ferrara, via Giuseppe Saragat 1, 44122 Ferrara, Italy}
\affiliation[c]{INAF/OAS Bologna, via Piero Gobetti 101, 40129 Bologna, Italy}
\affiliation[d]{Department of Physics \& Astronomy, University of the Western Cape, Cape Town 7535, South Africa}
\affiliation[e]{Instituto de Física Corpuscular (IFIC), CSIC‐Universitat de València, Parc Científic UV, c/ Catedrático José Beltrán, 2, E-46980 Paterna (València), Spain}
\affiliation[f]{Istituto Nazionale di Astrofisica, Osservatorio Astronomico di Trieste, via Tiepolo 11, 34143 Trieste, Italy}
\affiliation[g]{Institute for Fundamental Physics of the Universe, Via Beirut 2, 34151 Trieste, Italy}
\affiliation[h]{Istituto Nazionale di Fisica Nucleare, Sezione di Trieste, via Valerio 2, 34127 Trieste, Italy}
\affiliation[i]{Leinweber Institute for Theoretical Physics at Stanford, 382 Via Pueblo, Stanford, CA 94305, USA}
\affiliation[j]{Kavli Institute for Particle Astrophysics and Cosmology, 382 Via Pueblo, Stanford, CA 94305, USA}
\affiliation[k]{SISSA - International School for Advanced Studies, Via Bonomea 265, 34136 Trieste, Italy}
\affiliation[l]{Physik Department T31, Technische Universit\"at M\"unchen, James-Franck-Str.\ 1, D-85748 Garching, Germany}
\emailAdd{mario.ballardini@unife.it}
\emailAdd{barbieri@ific.uv.es}
\emailAdd{antonio.raffaelli@unife.it}
\emailAdd{chiara.moretti@inaf.it}
\emailAdd{ohep2@cantab.ac.uk}
\emailAdd{emiliano.sefusatti@inaf.it}
\emailAdd{francesco.verdiani@tum.de}

\abstract{
We search for primordial oscillatory features using cosmic microwave background (CMB) and large-scale structure (LSS) data, combining \textit{Planck} Public Release 4 (PR4) with galaxy clustering from the Baryon Oscillation Spectroscopic Survey (BOSS) and quasars from the extended Baryon Oscillation Spectroscopic Survey (eBOSS).

For the LSS data we perform a full-shape analysis of the power spectrum multipoles with the one-loop effective field theory of large-scale structure, applying infrared resummation separately to each oscillatory component. We consider four templates: linearly and logarithmically oscillating features in Fourier space, with constant, localised, or power-law-decaying amplitudes, and fit them to the CMB and LSS data separately and jointly.

We find no statistically significant evidence for primordial features in either dataset, individually or in combination. The best-fitting LSS features improve the fit by $\Delta\chi^2\simeq-6$ to $-10$, which corresponds to global significances of at most $1.6\,\sigma$ once the look-elsewhere effect is accounted for. In the CMB alone, the most significant feature is reduced to $\sim2.6\,\sigma$. Over the full frequency range, the LSS bounds on the feature amplitudes are weaker than those from \textit{Planck} by factors of $1.4$--$3$. Combining \textit{Planck} with \textit{BOSS+eBOSS} leaves the CMB constraints essentially unchanged, because the LSS data do not favour the frequencies preferred by the CMB.
}

\begin{document}

\maketitle

\section{Introduction}
The primordial power spectrum (PPS) of curvature perturbations encodes fundamental information about the dynamics of the early Universe. While the simplest single-field slow-roll inflationary models predict a nearly scale-invariant PPS~\cite{Starobinsky:1980te,Guth:1980zm,Linde:1981mu,Linde:1983gd,Mukhanov:1981xt}, deviations from this prediction, often referred to as \textit{primordial features}~\cite{Chen:2010xka,Chluba:2015bqa,Slosar:2019gvt,2022arXiv220308128A}, can arise due to sudden transitions in the inflationary potential~\cite{Starobinsky:1992ts,Adams:2001vc,Chen:2006xjb}, additional degrees of freedom during inflation~\cite{Achucarro:2010da,Chen:2011zf,Gao:2012uq,Chen:2014cwa,Braglia:2021rej}, modifications of the initial vacuum state of quantum fluctuations~\cite{Martin:2000xs,Easther:2002xe,Armendariz-Picon:2003knj}, or resonant effects~\cite{Silverstein:2008sg,Pahud:2008ae,Chen:2008wn,Flauger:2009ab,Flauger:2010ja,Chen:2010bka}.

Extensive searches for deviations from a power-law PPS have been carried out, primarily using data from the cosmic microwave background (CMB), but no statistically significant evidence has been found; see refs.~\cite{Adams:2001vc,Peiris:2003ff,Martin:2003sg,Covi:2006ci,Hamann:2007pa,Meerburg:2011gd,Planck:2013jfk,Meerburg:2013dla,Benetti:2013cja,Miranda:2013wxa,Easther:2013kla,Chen:2014joa,Achucarro:2014msa,Hazra:2014goa,Hazra:2014jwa,Hu:2014hra,Ade:2015lrj,Gruppuso:2015zia,Gruppuso:2015xqa,Hazra:2016fkm,Torrado:2016sls,Akrami:2018odb,Handley:2019fll,Zeng:2018ufm,Beutler:2019ojk,Canas-Herrera:2020mme,Braglia:2021ckn,Braglia:2021sun,Braglia:2021rej,Naik:2022mxn,Hamann:2021eyw,Ballardini:2022wzu,Antony:2022ert,Antony:2024vrx,Raffaelli:2025kew,Calderon:2025xod,AtacamaCosmologyTelescope:2025nti,Raffaelli:2026bhk,Nerval:2026iev,Hannestad:2026olh}. Current CMB constraints from \textit{Planck},  Atacama Cosmology Telescope and the South Pole Telescope data suggest that any deviations from scale invariance must be within a few percent of the PPS amplitude over the range $0.005 \lesssim k/(h\,{\rm Mpc}^{-1}) \lesssim 0.25$; see refs.~\cite{Raffaelli:2025kew,Cheslog:2026xyr}. A comprehensive reanalysis of the full \textit{Planck} temperature and polarisation data across independent high-$\ell$ likelihood pipelines and data releases is presented in the companion paper Raffaelli et al.~\cite{Raffaelli:2026bhk}: several previously reported hints of oscillatory structure are found to persist qualitatively across the different analyses, yet none reaches global statistical significance, with the most significant feature reduced to the $\sim 2.6\,\sigma$ level once the look-elsewhere effect (LEE) is accounted for.

Increasing attention has been paid to large-scale structure (LSS) of the Universe as a complementary way to test inflationary physics, see refs.~\cite{Wang:1998gb,Zhan:2005rz,Huang:2012mr,Chen:2016vvw,Chen:2016zuu,Ballardini:2016hpi,Xu:2016kwz,Fard:2017oex,Palma:2017wxu,Ballardini:2017qwq,Zeng:2018ufm,Ballardini:2018noo,Slosar:2019gvt,Beutler:2019ojk,Ballardini:2019tuc,Debono:2020emh,Li:2021jvz,Ballardini:2022wzu,Mergulhao:2023ukp,Euclid:2023shr,Ballardini:2024dto,Calderon:2025xod,Stahl:2025qru,Cohen:2026pat,Karagiannis:2026xkv,Chen:2026gxp}. The potential of LSS to probe primordial features lies in the opportunity to measure the power spectrum of matter perturbations over a wide range of scales, with a primordial signal that is not diluted by projection along the line of sight, unlike in the CMB angular power spectra, where oscillatory features in $k$ get smeared when projected onto multipoles $\ell$. In particular, spectroscopic galaxy surveys provide access to small-scale information where high-frequency oscillation patterns could be better constrained. However, using LSS data for this purpose requires an accurate theoretical framework capable of modelling nonlinear structure formation.

Describing the galaxy power spectrum on mildly nonlinear scales requires perturbative techniques to handle nonlinear gravitational evolution. Primordial oscillatory features have previously been incorporated into perturbation-theory frameworks built for baryon acoustic oscillations (BAO), in particular through infrared (IR) resummation methods developed to account for large-scale bulk flows~\cite{Crocce:2007dt,Creminelli:2013poa,Baldauf:2015xfa,Blas:2015qsi,Blas:2016sfa,Senatore:2017pbn,Ivanov:2018gjr,Vlah:2018ygt}. Applied to both linear and logarithmic oscillations~\cite{Blas:2016sfa,Vasudevan:2019ewf,Beutler:2019ojk,Chen:2020ckc,Ballardini:2024dto}, this approach provided valuable insight and was later exploited systematically in full-shape analyses of the galaxy power spectrum.
Independently, the effective field theory of large-scale structure (EFTofLSS)~\cite{Carrasco:2012cv,Baumann:2010tm} established a systematic, well-controlled description of the mildly nonlinear galaxy power spectrum, self-consistently incorporating bias and counterterm operators at fixed perturbative order. Full-shape analyses built on this framework have since become a standard tool for extracting cosmological information from spectroscopic galaxy surveys~\cite{Ivanov:2019pdj,DAmico:2019fhj,Ivanov:2019hqk,Nishimichi:2020tvu,Chudaykin:2020ghx,DESI:2024jxi,Chudaykin:2025aux}.
Only recently has the full redshift-space EFTofLSS description of the galaxy power spectrum, and in one case the bispectrum, been applied to primordial oscillatory features; see refs.~\cite{Calderon:2025xod,Chen:2026gxp} and this work.

We present a full-shape analysis of primordial oscillatory features using spectroscopic galaxy clustering data, first from Baryon Oscillation Spectroscopic Survey (BOSS)~\cite{BOSS:2016wmc} alone, then in combination with the extended Baryon Oscillation Spectroscopic Survey (eBOSS) \cite{eBOSS:2020uxp}, and finally jointly with \textit{Planck} CMB data. Our analysis differs from ref.~\cite{Calderon:2025xod} in the adopted IR-resummation scheme, based on time-sliced perturbation theory (TSPT) with an explicit wiggle/no-wiggle split (\cref{sec:eft}) rather than a resummation of the full linear power spectrum; relative to ref.~\cite{Chen:2026gxp} our analysis extends the search to a joint analysis with CMB data. Relative to both refs.~\cite{Calderon:2025xod,Chen:2026gxp}, we also consider two templates with modulated amplitude, and perform a dedicated sample-based search for local maxima in frequency, with cross-dataset and per-redshift-bin consistency checks (\cref{sec:results}). Our methodology builds on previous developments based on TSPT~\cite{Blas:2015qsi,Blas:2016sfa,Ivanov:2018gjr,Ballardini:2024dto} and aims to provide constraints on inflationary features beyond simplified BAO-based approximations as done in refs.~\cite{Beutler:2019ojk,Ballardini:2019tuc,Ballardini:2022wzu,Mergulhao:2023ukp}.

We structure the paper as follows. In~\cref{sec:theory}, we describe the feature templates and models considered in this analysis. The modelling of the galaxy power spectrum is described in~\cref{sec:eft} where also the EFTofLSS in the presence of primordial features is discussed.  Datasets and our analysis setup are presented in~\cref{sec:data}. In~\cref{sec:results}, we present the main results for LSS data alone and in combination with CMB data.
We finally conclude in~\cref{sec:conclusions}. 
\Cref{app:sigma_rsd} derives the redshift-space IR damping scales for the oscillatory templates. \Cref{app:cosmo_bias} collects posterior figures and summary table for the standard cosmological and bias parameters. \Cref{app:window} illustrates the effect of the survey window function and the validation of the pipeline on noise-free mock data in~\cref{app:asimov}. \Cref{app:integrals} validates the FFTLog integration of the one-loop power spectrum against brute-force numerical integration.

\section{Primordial scale-dependent oscillatory features} \label{sec:theory}
We consider four templates with oscillatory features on top of a standard, featureless power-law PPS. The scalar PPS is  parameterised as
\begin{equation} \label{eqn:prim_spectrum_split}
    P_\zeta(k) = P_{\zeta,0}(k) \left[1 + \delta P^X_\zeta(k)\right] \,,
\end{equation}
where $P_{\zeta, 0}$ is the standard power-law spectrum, given by
\begin{equation}
    P_{\zeta, 0}(k) \equiv \frac{2 \pi^2}{k^3} A_\mathrm{s} \left(\frac{k}{k_*}\right)^{n_\mathrm{s}-1} \,.
\end{equation}
Here, $A_\mathrm{s}$ is the scalar amplitude, $n_\mathrm{s}$ is the scalar spectral index, and $k_*$ is the pivot scale, fixed to $k_* = 0.05 \, \mathrm{Mpc}^{-1}$.
The term $\delta P^X_\zeta(k)$ represents the relative correction induced by the primordial features, for which we study four distinct physical scenarios.

Sharp, linear-in-$k$ oscillations arise when the inflaton undergoes a nearly instantaneous deviation from slow roll, such as a step or kink in the potential~\cite{Starobinsky:1992ts,Adams:2001vc,Chen:2006xjb}. In general, the resulting oscillation amplitude is not strictly constant: it follows an envelope set by the physical properties of the transition and both the envelope and phase depend on the inflationary potential and background evolution; in multifield scenarios, interactions with additional heavy fields further affect the amplitude and phase, introducing additional complexity~\cite{Achucarro:2010da,Gao:2012uq,Chen:2014cwa}. For a sufficiently sharp, quasi-instantaneous transition, however, this envelope varies slowly over the range of scales of interest and can be approximated as constant.
Logarithmic oscillations arise instead when the slow-roll potential experiences a periodic modulation, which translates into an oscillatory correction to the background evolution and, in turn, to the PPS~\cite{Chen:2008wn,Flauger:2009ab,Chen:2011zf}; in this case a constant amplitude holds generically. We refer to these two constant-amplitude templates, oscillating linearly and logarithmically in $k$ respectively, as the sharp (LIN) and resonant (LOG) models.

These two constant-amplitude features are conventionally parameterised as
\begin{equation} \label{eqn:Pk_template}
    \delta P_\zeta^X(k) = A_X \sin \left(\omega_X \Xi_X + \phi_X\right) \,,
\end{equation}
where $X \in \{\mathrm{lin}, \mathrm{log}\}$ for these two templates. The scale dependence of the phase argument is defined via the coordinate $\Xi_X = \{k/k_*, \, \ln(k/k_*)\}$. Here, $A_X$ denotes the dimensionless amplitude of the primordial features, $\omega_X$ is the dimensionless frequency, and $\phi_X$ is a phase shift.

Unlike the constant-amplitude templates above, a distinct class of scenarios produces localised oscillations with an intrinsically scale-dependent amplitude. Models with a step, dip, or bump in the inflationary potential induce local oscillations through a brief violation of the slow-roll conditions~\cite{Starobinsky:1992ts,Adams:2001vc}; sharp turns in the trajectory of multifield inflation~\cite{Achucarro:2010da} produce a similar effect, as do sudden variations in the speed of sound, which give rise to oscillatory signatures with a scale-dependent envelope~\cite{Bean:2008na}. While these scenarios differ in their detailed oscillation phase, envelope, and damping behaviour, we adopt as a representative template the bump/dip model (BUMP) computed in refs.~\cite{Braglia:2022ftm,Braglia:2026fle}, characterised by localised linear oscillations
\begin{equation} \label{eqn:BUMP_template}
    \delta P_\zeta^\mathrm{bump}(k) = A_\mathrm{bump} \sqrt{\frac{e}{2\alpha}} \left[2\cos(x_0) + \left(x_0 - \frac{2}{x_0}\right)\sin(x_0)\right] e^{-x_0^2/(4\alpha)} \,,
\end{equation}
where $x_0 \equiv k\,\omega_\mathrm{bump}/k_*$ maintains a consistent definition of the linear oscillations relative to~\cref{eqn:Pk_template}. The parameter $\alpha$ controls the duration of the slow-roll violation: a larger $\alpha$ corresponds to a shorter, sharper transient, while a smaller $\alpha$ describes a more extended one~\cite{Braglia:2022ftm}.

Finally, classical or quantum-mechanical oscillations of additional massive fields during inflation can act as standard clocks, directly imprinting the time dependence of the background scale factor onto the primordial spectrum~\cite{Chen:2011zf,Chen:2014cwa,Chen:2015lza}. We concentrate here on the classical regime, which induces logarithmic oscillations modulated by a scale-dependent envelope: the classical primordial standard clock model (PSC). The template takes the form
\begin{equation} \label{eqn:PSC_template}
    \delta P_\zeta^\mathrm{psc}(k) = 
    \begin{cases}
        0 \,, & \text{if } k < k_r \,,  \\
        A_\mathrm{psc} \left( \frac{k}{k_r}  \right)^{-\frac{3}{2}} \sin \left( \omega_\mathrm{psc} \ln \frac{k}{k_r} + \phi_\mathrm{psc} \right) \,,  & \text{if } k \geq k_r \,,
\end{cases}
\end{equation}
where $k_r$ represents the sharp turn or particle production scale triggering the clock signals, and the $k^{-3/2}$ envelope captures the characteristic decay of the massive field's oscillations on an inflationary background.
We stress that~\cref{eqn:PSC_template} only captures the resonant, late-time tail of the signal: a complete, model-dependent treatment generically also produces a sharp-feature component at the turn itself, which this phenomenological template does not attempt to capture~\cite{Braglia:2021ckn}.

\section{EFTofLSS galaxy power spectrum} \label{sec:eft}
We describe the nonlinear galaxy power spectrum in redshift space using the EFTofLSS. This approach systematically incorporates the impact of unresolved small-scale dynamics on mildly nonlinear scales by introducing controlled, scale-dependent corrections to the theoretical model.

\subsection{The matter power spectrum with primordial features in real space}
The information in the scalar PPS is transferred to the linear matter power spectrum as
\begin{equation} \label{eqn:lin_evol}
    P_L(k) = \left[ T(\eta_0, k) \right]^2 P_\zeta( k ) \,,
\end{equation}
where the transfer function, $T(\eta_0,k)$, encodes the physics of recombination, expansion and the growth of fluctuations at current conformal time $\eta_0$.  
We also split the matter power spectrum, decomposing the linear power spectrum into a smooth (non-wiggly, nw) and an oscillatory (wiggly, w) part as
\begin{equation} \label{eqn:matt_ps_split}
    P_L(k) = P_L^\mathrm{nw}(k) + P_L^\mathrm{w}(k) \,,
\end{equation}
where generically $P_L^\mathrm{w}$ accounts for the BAO features in a spatially flat $\Lambda$CDM cosmological model.
In the presence of primordial features, it is useful to further separate the wiggly components of the linear matter power spectrum as 
\begin{equation} \label{eqn:wiggle_spectrum}
    P^\mathrm{w}_L(k) = P_\mathrm{BAO}^\text{w}(k) + P_X^\mathrm{w}(k) + P_\mathrm{BAO}^\mathrm{w}(k) \delta P^X_\zeta(k) \,,
\end{equation}
where $P_\mathrm{BAO}^\mathrm{w}(k)$ is the standard BAO feature~\cite{Hu:1995en,Eisenstein:1997ik,Seo:2007ns}. 
The component proportional to the primordial features is given by 
\begin{equation}
    P_X^\mathrm{w}(k) = P^\mathrm{nw}_L(k) \delta P^X_\zeta(k) \,.
\end{equation}
Finally, the third term on the right-hand side in~\cref{eqn:wiggle_spectrum} is the cross term between the BAO and the primordial feature. 
Since it is proportional to $A_\mathrm{BAO}\, A_X$,\footnote{$A_\mathrm{BAO}$ denotes the amplitude of the BAO wiggles relative to $P^\mathrm{nw}_L$, $A_\mathrm{BAO}\lesssim0.1$ on the scales of interest.} it is suppressed by about an order of magnitude with respect to the feature term and oscillates at the sum and difference of the two frequencies; following ref.~\cite{Ballardini:2024dto}, where this term is studied in detail, we neglect it and do not include it in our model.

At one-loop order, standard perturbation theory (SPT) predicts that the matter power spectrum has the following form
\begin{equation}
    P_m(k) = P_L(k) + P^\text{1-loop}(k) \,,
\end{equation}
where
\begin{equation}
    P^\text{1-loop}(k) = P_{22}^\text{1-loop}(k) + P_{13}^\text{1-loop}(k) \,,
\end{equation}
and the one-loop corrections are given by
\begin{align}\label{eqn:Pspt}
    &P_{22}^\text{1-loop}(k) = 2 \int \frac{\dd^3 q}{(2\pi)^3} F_2^2(\mathbf{q},\mathbf{k-q}) P_L(q) P_L(\left|\mathbf{k-q}\right|) \,,\\
    &P_{13}^\text{1-loop}(k) = 6 P_L(k) \int \frac{\dd^3 q}{(2\pi)^3} F_3(\mathbf{q},-\mathbf{q},\mathbf{k}) P_L(q) \,,
\end{align}
with \(F_2\) and \(F_3\) the standard time-independent matter SPT kernels, which are quadratic and cubic, computed in the Einstein--de Sitter approximation; see ref.~\cite{Bernardeau:2001qr}.
To correct for the unphysical sensitivity of SPT to ultraviolet (UV) modes, counterterms must be introduced, capturing the departures from Eulerian hydrodynamics~\cite{Carrasco:2012cv}. In EFTofLSS, the counterterm at one-loop is given by
\begin{equation} \label{eqn:ctr}
    P_\mathrm{ctr}(k) = -2 \tilde{c}_0 k^2 P_L(k) \,,
\end{equation}
where $\tilde{c}_0$ is an effective parameter (which absorbs the dependence of $P_m$ on the maximum $q$ used in eq.~\ref{eqn:Pspt}).

At low redshifts, bulk flows damp the BAO feature in the power spectrum~\cite{Eisenstein:2006nj}, but their physical effect can be resummed non-perturbatively to all orders in perturbation theory~\cite{Senatore:2014via,Vlah:2015zda,Blas:2015qsi}, a procedure known as IR resummation and closely linked to the equivalence principle~\cite{Baldauf:2015xfa}. We base our study on TSPT~\cite{Blas:2015qsi,Blas:2016sfa}, a formulation free from the spurious IR divergences. 

At leading order (LO), the result of the IR-resummation is a damping factor applied to the wiggly contribution of the linear matter power spectrum 
\begin{equation} \label{eqn:Pm_lo}
    P_m^\mathrm{LO}(\eta, k) = g^2(\eta) \left[ P_L^\mathrm{nw}(k) + e^{-g^2(\eta) \mathcal{S}} P_L^\mathrm{w}(k)\right] \,,
\end{equation}
where $g(\eta)$ is the linear growth factor at conformal time $\eta$. 
The resummation procedure can be extended to include the corrections from hard loops~\cite{Blas:2016sfa}. For the next-to-leading order (NLO) in the hard loops, the result is 
\begin{align} \label{eqn:Pm_nlo}
    P_m^\mathrm{NLO}(\eta, k) &= g^2(\eta) \left[P_L^\mathrm{nw} \round{k} + \left(1 + g^2(\eta) \mathcal{S}\right) e^{-g^2(\eta) \mathcal{S} } P_L^\mathrm{w}(k)\right] \notag \\
    & \quad + g^4(\eta) P^\text{1-loop} \left[P_L^\mathrm{nw}(k) + e^{-g^2(\eta) \mathcal{S}} P_L^\mathrm{w}(k)\right] \,,
\end{align}
where the last term is the usual one-loop correction to the power spectrum evaluated using the linear spectrum with damped 
oscillations corresponding to the LO result. 

The \textit{damping operator} $\mathcal{S}$ is defined as
\begin{equation} \label{eqn:s_operator}
    \mathcal{S} \equiv - \frac{1}{2} \int_{q \leq k_\mathrm{S}} \frac{\dd^3 q}{(2\pi)^3} \, P_L^\mathrm{nw} \round{q} \mathcal{D}_\mathbf{q} \mathcal{D}_{-\mathbf{q}} \,,
\end{equation}
 where $k_\mathrm{S} = 0.2\,h\,\text{Mpc}^{-1}$ is the separation scale between soft and hard momenta, and the \textit{derivative operator} $\mathcal{D}_\mathbf{q}$ acting on the wiggly power spectrum is
\begin{align} 
    \mathcal{D}_\mathbf{q} P_L^\mathrm{w} \round{k} &\equiv \frac{\round{\mathbf{q} \cdot \mathbf{k}}}{q^2} \sround{P_L^\mathrm{w} \round{\abs{\mathbf{k} + \mathbf{q}}} - P_L^\mathrm{w} \round{k}} \notag\\
    &= \left. \frac{\round{\mathbf{q} \cdot \mathbf{k}}}{q^2} \sround{e^{\mathbf{q} \cdot \nabla_{\mathbf{k'}}} -1} P_L^\mathrm{w} \round{k'} \right\rvert_{\mathbf{k}' = \mathbf{k}} \,. \label{eqn:d_op_def}
\end{align}
The action of the damping operator on the wiggly part of the spectrum is
\begin{align} 
    \mathcal{S} P_L^\mathrm{w} \left(k\right) 
    &= \left. \int_{q \leq k_\mathrm{S}} \frac{\dd^3 q}{\left(2 \pi\right)^3} \; P_L^\mathrm{nw} \left(q\right) \frac{\left(\mathbf{q} \cdot \mathbf{k}'\right)^2}{q^4} \left[ 1 - \cosh \left( \mathbf{q} \cdot \nabla_{\mathbf{k'}} \right) \right] P_L^\mathrm{w} \left(k'\right) \right\rvert_{\mathbf{k}'= \mathbf{k}} \,,\notag\\
    &= \left. \int_{q \leq k_\mathrm{S}} \frac{\dd^3 q}{\left(2 \pi\right)^3} \; P_L^\mathrm{nw} \left(q\right) \frac{\left(\mathbf{q} \cdot \mathbf{k}'\right)^2}{q^4} \left[ 1- \cos \left(\tilde{\omega}_X \mathbf{q} \cdot \mathbf{\hat{k}'} \right) \right] P_L^\mathrm{w}(k') \right\rvert_{\mathbf{k}'= \mathbf{k}} \,, \notag\\
    &= P_L^\mathrm{w}(k) \frac{k^2}{6 \pi^2} \int_0^{k_{\rm S}} \dd q \; P_L^\mathrm{nw}(q) \left[ 1 - j_0 \left(\tilde{\omega}_X q\right) + 2 j_2 \left(\tilde{\omega}_X q\right) \right] \notag  \\ 
    &\equiv k^2 \Sigma_X \left(k \right) P_L^\mathrm{w}(k) \,, \label{eqn:s_osc}
\end{align}
where $\tilde{\omega}_X$ is the dimensional frequency of the wiggly component $X$: $\tilde{\omega}_\mathrm{BAO} \simeq 110 \,h^{-1}\,\mathrm{Mpc}$, $\tilde{\omega}_\mathrm{lin} \equiv \omega_\mathrm{lin} /k_*$, and $\tilde{\omega}_\mathrm{log} \equiv \omega_\mathrm{log}/k$. \Cref{eqn:s_osc} applies separately to each oscillatory component of~\cref{eqn:wiggle_spectrum}, with $\tilde\omega$ set by the frequency of the specific wiggly piece being resummed: for the LIN template, whose phase is linear in $k$, $\tilde\omega_\mathrm{lin}$ is a genuine constant, as is $\tilde\omega_\mathrm{BAO}$; for BUMP, $\tilde\omega_\mathrm{bump}\equiv\omega_\mathrm{bump}/k_*$ is also constant, but the envelope varies on scales $\Delta k\simeq k_*\sqrt{2\alpha}/\omega_\mathrm{bump}$, so~\cref{eqn:s_osc} holds at leading order. For the LOG and PSC templates, whose phase runs as $\ln k$, the local frequency $\tilde\omega_\mathrm{log}(k)$ (and, for $k\geq k_r$, $\tilde\omega_\mathrm{psc} = \omega_\mathrm{psc}/k$) depends on the external wavenumber, and we evaluate it at $k$. Expanding the phase, $\omega_\mathrm{log}\ln(1+q\nu/k)=\tilde\omega_\mathrm{log}q\nu-\omega_\mathrm{log}q^2\nu^2/(2k^2)+\dots$, so freezing the frequency is formally justified only for $\omega_\mathrm{log}q^2/k^2\ll1$. The approximation is nonetheless adequate in practice: at low frequency this condition holds over the soft-loop range, while at high frequency $\tilde\omega_X q\gg1$ and the kernel $1-j_0+2j_2$ saturates to unity, so that $\Sigma_X$ becomes insensitive to the precise value of the frequency (see also~\cref{app:log_exact}).

\subsection{The galaxy power spectrum with primordial features in real space}
Having computed the IR-resummed matter power spectrum at one-loop in real space, we turn now to the galaxy power spectrum, starting in real space. The galaxy distribution is biased with respect to the matter one, so we need to model this biasing through a perturbative expansion; see refs.~\cite{Chan:2012jj,Baldauf:2012hs,Abidi:2018eyd,Perko:2016puo}; see ref.~\cite{Desjacques:2016bnm} for a review. The bias expansion up to third order can be written as
\begin{align} \label{eqn:delta_g_third}
    \delta_g=\, & b_{1} \delta + \frac{b_{2}}{2} \delta^{2} + \frac{b_{3}}{6} \delta^{3} \notag\\
    &+ b_{\mathcal{G}_2} \mathcal{G}_2 + b_{\mathcal{G}_3} \mathcal{G}_3 + b_{\mathcal{G}_2 \delta} \mathcal{G}_2 \delta + b_{\Gamma_{3}}\Gamma_{3} \notag\\
    &+ \epsilon + \epsilon_{\delta} \delta + \epsilon_{\delta^2} \delta^2 + \epsilon_{\mathcal{G}_2} \mathcal{G}_2 \notag\\
    &+ b_{\nabla^2\delta} \nabla^2\delta \,,
\end{align}
where so-called local bias operators such as $\delta, \delta^{2}$, and $\delta^{3}$ and non-local operators $\mathcal{G}_{2}, \mathcal{G}_{3}$ and $\Gamma_{3}$ have been introduced as all the operators allowed by symmetry at this order. The latter are defined in terms of the density and velocity divergence potentials $\Phi$ and $\Phi_{v}$ as
\begin{align}
\mathcal{G}_{2} & \equiv\left[\left(\partial_{i} \partial_{j} \Phi\right)^{2}-\left(\nabla^{2} \Phi\right)^{2}\right] \,, \\
\mathcal{G}_{3} & \equiv -\frac{1}{2} \left[2\left(\partial_i\partial_j\Phi\right)^3 + \delta^3 - 3\left(\partial_i\partial_j\Phi\right)^2\delta\right] \,, \\
\Gamma_{3} & \equiv\left(\partial_{i} \partial_{j} \Phi\right)^{2}-\left(\nabla^{2} \Phi\right)^{2}-\left[\left(\partial_{i} \partial_{j} \Phi_{v}\right)^{2}-\left(\nabla^{2} \Phi_{v}\right)^{2}\right] \,.
\end{align}
The expression in~\cref{eqn:delta_g_third} includes also the stochastic contributions $\epsilon, \epsilon_{\delta} \delta$, $\epsilon_{\delta^{2}} \delta^{2}$, $\epsilon_{\mathcal{G}_2}$ and the higher-derivative operator $\nabla^{2} \delta$.

Following refs.~\cite{Desjacques:2017msa,Abidi:2018eyd,Ivanov:2018gjr}, the third-order operators $b_3\delta^3$, $b_{\mathcal{G}_3}\mathcal{G}_3$, and $b_{\mathcal{G}_2\delta}\mathcal{G}_2\delta$ are removed by bias renormalisation \citep{Assassi:2014fva}, and we only keep these terms
\begin{equation} \label{eqn:delta_g_third_2}
    \delta_g = b_{1} \delta + \frac{b_{2}}{2} \delta^{2} + b_{\mathcal{G}_2} \mathcal{G}_2 + b_{\Gamma_{3}}\Gamma_{3} + \epsilon  \,.
\end{equation}
Given this bias expansion, the model for the one-loop galaxy power spectrum in real space in the EFTofLSS is
\begin{equation}
    P_g(k) = K_1^2 P_L(k) + P^\text{1-loop}_g(k) + P_\mathrm{ctr}(k) + P_\mathrm{noise}(k) \,,
\end{equation}
where
\begin{align}
    &P^\text{1-loop}_g(k) = P_{g,22}^\text{1-loop}(k) + P_{g,13}^\text{1-loop}(k) \,,\\
    &P_{g,22}^\text{1-loop}(k) = 2 \int \frac{\dd^3 q}{(2\pi)^3} K_2^2(\mathbf{q},\mathbf{k-q}) P_L(q) P_L(\left|\mathbf{k-q}\right|) \,,\\
    &P_{g,13}^\text{1-loop}(k) = 6 K_1 P_L(k) \int \frac{\dd^3 q}{(2\pi)^3} K_3(\mathbf{q},-\mathbf{q},\mathbf{k}) P_L(q) \,,
\end{align}
where the $K_n$ kernels collect the nonlinear bias terms in real space, in the notation of ref.~\cite{Ivanov:2018gjr}, with $K_1 = b_1$.

Following the same IR-resummation procedure derived for the matter power spectrum, \cref{eqn:Pm_lo,eqn:Pm_nlo}, with the damping operator $\mathcal{S}$ defined as in~\cref{eqn:s_operator}, the IR-resummed galaxy power spectrum at LO and NLO in real space read
\begin{align}
    &P_g^\mathrm{LO}(\eta, k) = g^2(\eta) K_1^2 \left[ P_L^\mathrm{nw}(k) + e^{-g^2(\eta) \mathcal{S}} P_L^\mathrm{w}(k)\right] \,,\\
    &P_g^\mathrm{NLO}(\eta, k) = g^2(\eta) K_1^2 \left[P_L^\mathrm{nw}(k) + \left(1 + g^2(\eta) \mathcal{S}\right) e^{-g^2(\eta) \mathcal{S}} P_L^\mathrm{w}(k)\right] \notag \\
    & \quad + g^4(\eta) P_g^\text{1-loop} \left[P_L^\mathrm{nw}(k) + e^{-g^2(\eta) \mathcal{S}} P_L^\mathrm{w}(k)\right] \,.
\end{align}
The counterterm contribution takes the same form as in real space for the matter field,~\cref{eqn:ctr}, with $\tilde{c}_0$ now also absorbing the bias-dependent UV sensitivity of the galaxy field.

Finally, the shot-noise contribution to the galaxy power spectrum in real space, due to the auto-correlation of any galaxy with itself as well as from stochastic operators in the bias expansion, can be modelled at large scales as
\begin{equation}
    P_\mathrm{noise}(k) = \frac{1}{\bar{n}}\left(\alpha_P + \epsilon_0 k^2\right) \,,
\end{equation}
where the constant term $\alpha_P$ accounts for Poisson shot noise as well as potential deviations from it, while $\epsilon_0$ accounts for scale dependence; $\bar{n}$ is the mean comoving number density of the tracer.

\subsection{The galaxy power spectrum with primordial features in redshift space}
The calculation of the galaxy power spectrum in redshift space is crucial for interpreting observations from galaxy surveys, where peculiar velocities distort the clustering signal along the line of sight.
We adopt the plane-parallel (or flat-sky) approximation, valid when the separation between points is much smaller than their distance to the observer, under which the real-space coordinate $\mathbf{x}$ and the redshift-space coordinate $\mathbf{s}$ are related via Hubble's law by
\begin{equation}
    \mathbf{s} = \mathbf{x} + \hat{\mathbf{z}} \frac{v^{(r)}_z(\eta, \mathbf{x})}{\mathcal{H}} \,,
\end{equation}
where $\mathcal{H}$ is the conformal Hubble parameter, $\hat{\mathbf{z}}$ the unit vector along the line of sight, and $v^{(r)}_z(\eta,\mathbf{x})$ the line-of-sight component of the peculiar velocity. We denote real-space quantities with a superscript $(r)$ and their redshift-space counterparts with $(s)$.

At linear order, these distortions are described by the Kaiser effect~\cite{Kaiser:1987qv}, which enhances the power spectrum anisotropically, introducing a dependence on the angle between the wave vector $\mathbf{k}$ and the line-of-sight direction $\hat{\mathbf{z}}$ as
\begin{equation}
    P^{(s)}_L(k,\mu) = Z_1^2(\mu) P_L(k) \,,
\end{equation}
where $\mu \equiv \hat{\mathbf{k}}\cdot\hat{\mathbf{z}}$, $Z_1(\mu) = b_1 + f\mu^2$, and $f$ is the growth rate of matter perturbations, which we assume to be scale independent; the redshift-space matter power spectrum is recovered for $b_1=1$, and the real-space results of~\cref{eqn:Pm_lo,eqn:Pm_nlo} for $b_1=1$ and $f=0$.

At one-loop order, the nonlinear coupling of density, velocity and bias fields introduces scale-dependent modifications to the redshift-space galaxy power spectrum, including density-density, density-velocity, and velocity-velocity contributions that drive the damping and mode coupling observed at smaller scales
\begin{equation} \label{eqn:P_g_redshift}
    P^{(s)}_\mathrm{g}(k,\mu) = Z_1^2(\mu) P_L(k) + P^{(s)\, \text{1-loop}}_g(k,\mu) + P^{(s)}_\text{ctr}(k,\mu) + P^{(s)}_\text{noise}(k,\mu) \,,
\end{equation}
with
\begin{align}
    &P_{g,22}^{(s)\, \text{1-loop}}(k,\mu) = 2 \int \frac{\dd^3 q}{(2\pi)^3} Z_2^2(\mathbf{q},\mathbf{k-q}) P_L(q) P_L(\left|\mathbf{k-q}\right|) \,,\\
    &P_{g,13}^{(s)\, \text{1-loop}}(k,\mu) = 6 Z_1(\mu) P_L(k) \int \frac{\dd^3 q}{(2\pi)^3} Z_3(\mathbf{q},-\mathbf{q},\mathbf{k}) P_L(q) \,,
\end{align}
where the $Z_n$ kernels collect the nonlinear bias and velocity terms in redshift space, in the notation of ref.~\cite{Ivanov:2018gjr}, built from the real-space bias kernels $K_n$ introduced above together with the velocity/RSD contributions.

The IR resummation extends analogously to redshift space. The action of the operator $\mathcal{S}$ on the wiggly part of the power spectrum~\cref{eqn:s_osc} becomes, for a constant frequency $\tilde\omega$,
\begin{equation} \label{eqn:S_lin_rsd}
    \left.\mathcal{S}^{(s)} P_L^\mathrm{w} \left(k\right) = \mathcal{P}_{ab} \mathcal{P}_{cd} k'^a k'^c \int_{q\leq k_\mathrm{S}} \frac{\dd^3 q}{\left(2 \pi\right)^3} \, P_L^\mathrm{nw} \frac{q^b q^d}{q^4}
\left[1 - \cos \left(\tilde{\omega} \mathbf{q} \cdot \hat{\mathbf{k'}}\right) \right] P_L^\mathrm{w} \left(k'\right) \right\rvert_{\mathbf{k}' = \mathbf{k}} \,,
\end{equation}
where we define the projector tensor as $\mathcal{P}_{ab} \equiv \delta_{ab}^{\rm K} + f \hat{z}_a \hat{z}_b$. In real space we have only the Kronecker symbol $\delta_{ab}^{\rm K}$.
This follows from introducing the linear operator \( \mathcal{D}^{(s)}_\mathbf{q} \) in redshift space acting on the wiggly power spectrum as
\begin{align}
    \mathcal{D}^{(s)}_\mathbf{q} P_L^\mathrm{w}(k) &\equiv \frac{\mathcal{P}_{ab} k^a q^b}{q^2} \left[P_L^\mathrm{w}(|\mathbf{k} + \mathbf{q}|) - P_L^\mathrm{w}(k)\right] \,,\\
    &= \frac{\mathcal{P}_{ab} k^a q^b}{q^2} \left(e^{\mathbf{q} \cdot \nabla_\mathbf{k'}} - 1\right) P_L^\mathrm{w}(k') \Big|_{\mathbf{k'} = \mathbf{k}} \,. \label{eqn:d_op_def_rsd}
\end{align}
Evaluating this expression in an explicit coordinate system and integrating over the angles (\cref{app:sigma_rsd}) yields
\begin{equation}
    \mathcal{S}^{(s)} P_L^\mathrm{w} \left(k\right) \equiv k^2 \Sigma^{(s)}_X(k,\mu) P_L^\mathrm{w} \left(k\right) \,, \label{eqn:Sigma_s_def}
\end{equation}
where $\Sigma^{(s)}_X(k,\mu)$ takes the same anisotropic parallel/perpendicular form weighted by the Kaiser factor $\mu^2$, as the redshift-space IR resummation damping factor used for the BAO signal~\cite{Ivanov:2018gjr} and, by extension, for the feature templates. The explicit forms are given in~\cref{app:sigma_rsd}.

As in real space, this closed form is the standard leading-order result for the BAO, LIN and BUMP contributions to $P_L^\mathrm{w}(k)$, for which $\tilde\omega_X$ is constant, and an approximation for the LOG and PSC contributions once $\tilde\omega_X(k)$ is evaluated locally at the external wavenumber $k$, as discussed below~\cref{eqn:s_osc}; the treatment for the LOG template beyond the soft-loop-momentum approximation is given in~\cref{app:log_exact}.
The IR-resummed galaxy power spectrum at LO and NLO in redshift space read
\begin{align}
    &P_g^{(s)\, \mathrm{LO}}(\eta, k, \mu) = g^2(\eta) Z_1^2(\mu) \left[ P_L^\mathrm{nw}(k) + e^{-g^2(\eta) \mathcal{S}^{(s)}} P_L^\mathrm{w}(k)\right] \,,\\
    &P_g^{(s)\, \mathrm{NLO}}(\eta, k, \mu) = g^2(\eta) Z_1^2(\mu) \left[P_L^\mathrm{nw}(k) + \left(1 + g^2(\eta) \mathcal{S}^{(s)}\right) e^{-g^2(\eta) \mathcal{S}^{(s)}} P_L^\mathrm{w}(k)\right] \notag \\
    & \quad + g^4(\eta) P_g^{(s)\, \text{1-loop}} \left[P_L^\mathrm{nw}(k) + e^{-g^2(\eta) \mathcal{S}^{(s)}} P_L^\mathrm{w}(k)\right] \,.
\end{align}

In redshift space, the counterterm~\cref{eqn:ctr} contribution generalises to
\begin{align} \label{eqn:ctr_rsd}
    P_\text{ctr}^{(s)}(k,\mu) &= -2 k^2 \left(\tilde{c}_0 + \tilde{c}_2 f\mu^2 + \tilde{c}_4 f^2\mu^4\right) P_L(k) \notag\\
    &\quad + c_{\nabla^4 \delta} k^4 f^4 \mu^4 Z_1^2(\mu) P_L(k) \,,
\end{align}
where the first line corresponds to the generalisation of~\cref{eqn:ctr} to redshift space and the second line arises from higher-order contributions (formally arising at two-loop order) due to nonlinear RSD (Fingers-of-God effects).
Following ref.~\cite{Carrilho:2022mon}, we redefine the counterterms as
\begin{equation}
    c_0 = \tilde{c}_0 + \frac{f}{3} \tilde{c}_2 + \frac{f^2}{5}\tilde{c}_4\,,\quad c_2=\tilde{c}_2 + \frac{6f}{7}\tilde{c}_4,\,\quad c_4=\tilde{c}_4\,.
\end{equation}
We set $\tilde c_4=0$, since only $P_0$ and $P_2$ are fitted.

Practically, up to errors of the order of the NLO soft corrections, we can rewrite
\begin{align}
    &P_g^{(s)\, \mathrm{NLO}}(\eta, k, \mu) \simeq g^2(\eta) Z_1^2(\mu) \left[P_L^\mathrm{nw}(k) + \left(1 + g^2(\eta) \mathcal{S}^{(s)}\right) e^{-g^2(\eta) \mathcal{S}^{(s)}} P_L^\mathrm{w}(k)\right] \notag \\
    & \quad + g^4(\eta) \left[P_g^{(s)\, \text{1-loop}} \left[P_L^\mathrm{nw}(k)\right] + e^{-g^2(\eta) \mathcal{S}^{(s)}} P_g^{(s)\, \text{1-loop}} \left[P_L^\mathrm{w}(k)\right] \right] \,,
\end{align}
where $P_g^{(s)\, \text{1-loop}} \left[P_L^\mathrm{nw}(k)\right]$ is evaluated on the smooth power spectrum only and the one-loop component of the wiggly part is calculated as
\begin{equation}
    P_g^{(s)\, \text{1-loop}} \left[P_L^\mathrm{w}(k)\right] \simeq P_g^{(s)\, \text{1-loop}} \left[P_L(k)\right] - P_g^{(s)\, \text{1-loop}} \left[P_L^\mathrm{nw}(k)\right] \,.
\end{equation}

Finally, we also introduce the shot-noise contribution to the galaxy power spectrum, due to the auto-correlation of any galaxy with itself as well as from stochastic operators in the bias expansion. At large scales, this term can be modelled as
\begin{equation} \label{eqn:noise}
    P^{(s)}_\text{noise}(k,\mu) = \frac{1}{\bar{n}}\left(\alpha_P + \epsilon_0 k^2 + \epsilon_2 k^2 \mu^2\right) \,,
\end{equation}
where $\epsilon_2$ account for scale dependence with distinct redshift-space behaviour.

\subsection{Alcock--Paczynski effect and multipole projection}
Converting measured redshifts and angular positions into comoving wavenumbers and angles requires assuming a fiducial cosmology. A mismatch between this fiducial cosmology and the true one introduces an anisotropic geometrical distortion, the Alcock--Paczynski (AP) effect~\cite{Alcock:1979mp}, which we account for following the standard full-shape treatment~\cite{Ivanov:2018gjr,Ivanov:2019pdj,Beutler:2019ojk}. Defining the dilation parameters
\begin{equation}
    \alpha_\parallel \equiv \frac{H^\mathrm{fid}(z)}{H(z)} \,, \qquad \alpha_\perp \equiv \frac{D_A(z)}{D_A^\mathrm{fid}(z)} \,, \qquad F \equiv \frac{\alpha_\parallel}{\alpha_\perp} \,,
\end{equation}
where $H(z)$ and $D_A(z)$ are the true Hubble rate and angular diameter distance and the superscript ``fid'' denotes their fiducial-cosmology counterpart, the wavenumber and line-of-sight angle inferred in the fiducial cosmology, $k_\mathrm{obs}$ and $\mu_\mathrm{obs}$, relate to their true values $(k,\mu)$ entering~\cref{eqn:P_g_redshift,eqn:ctr_rsd,eqn:noise} as
\begin{equation}
    k = \frac{k_\mathrm{obs}}{\alpha_\perp}\left[1+\mu_\mathrm{obs}^2\left(\frac{1}{F^2}-1\right)\right]^{1/2} \,, \qquad
    \mu = \frac{\mu_\mathrm{obs}}{F}\left[1+\mu_\mathrm{obs}^2\left(\frac{1}{F^2}-1\right)\right]^{-1/2} \,.
\end{equation}
The associated Jacobian rescales the observed anisotropic power spectrum as
\begin{equation}
    P_g^{(s)\,\mathrm{obs}}(k_\mathrm{obs},\mu_\mathrm{obs}) = \frac{1}{\alpha_\parallel \alpha_\perp^2} P_g^{(s)}\left[k(k_\mathrm{obs},\mu_\mathrm{obs}),\mu(\mu_\mathrm{obs})\right] \,.
\end{equation}

Finally, we project the AP-corrected anisotropic galaxy power spectrum onto multipoles as
\begin{equation} \label{eqn:multipoles}
    P_{g,\ell}^{(s)}(k_\mathrm{obs}) = \frac{2\ell + 1}{2} \int_{-1}^1 \dd\mu_\mathrm{obs}\, P^{(s)\,\mathrm{obs}}_\mathrm{g}(k_\mathrm{obs},\mu_\mathrm{obs}) \, \mathcal{L}_\ell(\mu_\mathrm{obs}) \,,
\end{equation}
where $\mathcal{L}_\ell(\mu)$ are the Legendre polynomials.

\section{Cosmological datasets and analysis pipeline} \label{sec:data}
This section describes the datasets, numerical tools and settings used to perform the statistical analysis and parameter inference.

\subsection{CMB data} \label{sec:cmb_data}
For our baseline CMB likelihood configuration, we base the analysis on the following combination of \textit{Planck} likelihoods: \texttt{Commander} for large-scale temperature anisotropies ($\ell < 30$) \cite{Planck_likelihood_2020}, \texttt{SRoll2} for low-$\ell$ E-mode polarisation data \cite{Delouis_2019,Pagano:2019tci}, and for the high multipoles  ($\ell \geq 30$), the \texttt{CamSpec} temperature and polarisation likelihood ($TT, TE, EE$) based on the PR4 maps produced by the \texttt{NPIPE} processing pipeline \cite{Efstathiou_2021,camspec_npipe_2022}. All these likelihoods are unbinned in multipoles. We include the \textit{Planck} PR4 CMB lensing likelihood \cite{Carron_2022} reconstructed from the same maps.

A detailed performance comparison and consistency check between alternative \textit{Planck} high-$\ell$ formulations, specifically comparing \texttt{Plik} and \texttt{CamSpec} likelihoods for PR3 with \texttt{CamSpec} for PR4, has been performed in ref.~\cite{Raffaelli:2026bhk}. In this work, the CMB likelihood is added to the LSS analysis of~\cref{sec:lss_data} as an independent dataset; the cross-covariance between the CMB lensing and LSS likelihoods, expected to be small given their different redshift kernels, is neglected.

Parameter inference for the CMB-only and joint CMB+LSS analyses is carried out using the nested sampling algorithm implemented in \texttt{PolyChord}~\cite{Polychord1_Handley_2015,Polychord2_Handley_2015}, within the \texttt{Cobaya}~\cite{Cobaya_Torrado_2021} framework. For all runs, the number of live points is set to 1\,000 to ensure a robust sampling and accurate Bayesian evidence estimation. Theoretical predictions for the CMB angular power spectra and the linear matter power spectrum are computed using a modified version of the Boltzmann solver \texttt{CAMB}~\cite{CAMB_Lewis_2000}. The primordial helium abundance is consistently set using a compilation from the public code \texttt{PArthENoPE}~\cite{Gariazzo:2021iiu}.

All priors for the standard cosmological and primordial feature parameters are listed in~\cref{tab:priors_cmb,tab:priors_features}, respectively. 
The high-$\ell$ foreground and nuisance parameters associated with the \texttt{CamSpec} likelihood are all consistently varied according to ref.~\cite{camspec_npipe_2022}.
\begin{table}[t]
    \centering
    \begin{minipage}[t]{0.49\linewidth}
        \vspace{0pt}
        \centering
        \begin{tabular} { l c }
        \noalign{\vskip 3pt}\hline\noalign{\vskip 1.5pt}\hline\noalign{\vskip 5pt}
        \multicolumn{1}{c}{\bf Parameter} & \multicolumn{1}{c}{\bf Prior range}\\
        \noalign{\vskip 3pt}\cline{1-2}\noalign{\vskip 3pt}
        $\omega_\mathrm{c}$ & [0.095,\,0.145] \\
        $\omega_\mathrm{b}$ & [0.019,\,0.025] \\
        $\tau_\text{reio}$ & [0.01,\,0.3] \\
        $100\,\theta_\text{MC}$ & [1.03,\,1.05] \\\hline
        \noalign{\vskip 1.5pt}
        $\ln\left(10^{10}A_\mathrm{s}\right)$ & [1.61,\,3.91]\\
        $n_\mathrm{s}$ & [0.8,\,1.2]\\
        \hline\noalign{\vskip 1.5pt}\hline
        \end{tabular}
        \caption{Priors for the standard cosmological parameters used to analyse the CMB data.}
        \label{tab:priors_cmb}

        \vspace{0.5cm}

        \begin{tabular} { l c }
        \noalign{\vskip 3pt}\hline\noalign{\vskip 1.5pt}\hline\noalign{\vskip 5pt}
        \multicolumn{1}{c}{\bf Parameter} &
        \multicolumn{1}{c}{\shortstack{\bf Prior range /\\ \bf (mean, $\sigma$)}}\\
        \noalign{\vskip 3pt}\cline{1-2}\noalign{\vskip 3pt}
        $\omega_\mathrm{c}$ & [0.07,\,0.2] \\
        $\omega_\mathrm{b}$ & $(0.02268,\,0.00036)$ \\
        $h$ & [0.6,\,0.83] \\\hline
        \noalign{\vskip 1.5pt}
        $\ln\left(10^{10}A_\mathrm{s}\right)$ & [2,\,4]\\
        $n_\mathrm{s}$ & $(0.9649,\,0.0042)$\\
        \hline\noalign{\vskip 1.5pt}\hline
        \end{tabular}
        \caption{Priors for the standard cosmological parameters used to analyse the LSS data.}
        \label{tab:priors_eft}
    \end{minipage}
    \hfill
    \begin{minipage}[t]{0.49\linewidth}
        \vspace{0pt}
        \centering
        \begin{tabular} { l c }
        \noalign{\vskip 3pt}\hline\noalign{\vskip 1.5pt}\hline\noalign{\vskip 5pt}
        \multicolumn{1}{c}{\bf Parameter} & \multicolumn{1}{c}{\bf Prior range}\\
        \noalign{\vskip 3pt}\cline{1-2}\noalign{\vskip 3pt}
        $A_\mathrm{lin}$ & [0,\,0.5] \\
        $\log_{10} \omega_\mathrm{lin}$ & [0,\,2.1]$_\mathrm{CMB}$\\
           & [0.68,\,1.97]$_\mathrm{LSS}$\\
        $\phi_\mathrm{lin}/(2\pi)$ & [0,\,1]\\
        \noalign{\vskip 1.5pt}\hline
        $A_\mathrm{log}$ & [0,\,0.5] \\
        $\log_{10} \omega_\mathrm{log}$ & [0,\,2.1]$_\mathrm{CMB}$\\
           & [0.6,\,2.1]$_\mathrm{LSS}$\\
        $\phi_\mathrm{log}/(2\pi)$ & [0,\,1]\\
        \noalign{\vskip 1.5pt}\hline
        $A_\mathrm{bump}$ & [-0.5,\,0.5] \\
        $\log_{10} \omega_\mathrm{bump}$ & [0,\,2.1]$_\mathrm{CMB}$\\
           & [0.68,\,1.97]$_\mathrm{LSS}$\\
        $\log_{10} \alpha$ & [1.5,\,4]\\
        \hline\noalign{\vskip 1.5pt}\hline
        $A_\mathrm{psc}$ & [0,\,0.5] \\
        $\log_{10} \omega_\mathrm{psc}$ & [0,\,2.1]$_\mathrm{CMB}$\\
           & [0.6,\,2.1]$_\mathrm{LSS}$\\
        $\phi_\mathrm{psc}/(2\pi)$ & [0,\,1]\\
        $\log_{10} \left(k_r\,\text{Mpc}\right)$ & [-2.3,\,-0.8]$_\mathrm{CMB}$\\
           & [-2.0,\,-0.8]$_\mathrm{LSS}$\\
        \noalign{\vskip 1.5pt}\hline\hline
        \end{tabular}
        \caption{Priors on the feature parameters, for the CMB and the LSS datasets. The frequency bounds for the LSS analysis follow the criteria of~\cref{sec:freq_range}.}
        \label{tab:priors_features}
    \end{minipage}
\end{table}

\subsection{LSS data} \label{sec:lss_data}
For the LSS analysis, we use redshift-space galaxy power spectrum multipole measurements from the completed SDSS-III BOSS DR12 galaxy sample~\cite{BOSS:2016wmc} and, in combination, from the SDSS-IV eBOSS DR16 quasar sample~\cite{eBOSS:2020uxp}. We consider two galaxy-clustering dataset combinations, hereafter referred to as:
\begin{itemize}
    \item \textit{BOSS}: the BOSS DR12 LOWZ and CMASS North and South Galactic Cap catalogues, combined into two redshift bins ($z_\mathrm{eff} = 0.38$ and $0.61$). We use the power spectrum monopole and quadrupole ($P_0, P_2$) over the range $k \in [0.0025,\,0.20]\,h\,\mathrm{Mpc}^{-1}$, and combine the North and South Galactic Caps according to their relative areas as in ref.~\cite{Gil-Marin:2016wya}.
    \item \textit{BOSS+eBOSS}: the same two BOSS redshift bins combined with the eBOSS DR16 quasar sample in a third, higher-redshift bin ($z_\mathrm{eff} = 1.48$), for a total of three redshift bins fit simultaneously. The quasar power spectrum multipoles cover the range $k \in [0.02,\,0.30]\,h\,\mathrm{Mpc}^{-1}$, while the two BOSS bins retain the range above. 
\end{itemize}
All power spectra are measured with the \textsc{PolySpec} code of ref.~\citep{Philcox:2024rqr} (based on ref.~\cite{Philcox:2020vbm}), analogously to the Dark Energy Spectroscopic Instrument (DESI) power spectrum measurements described in ref.~\citep{Chudaykin:2025aux}. Each power spectrum multipole is measured in linearly spaced $k$-bins with $\Delta k=0.0025\,h\,\text{Mpc}^{-1}$, and we compute also a \textit{theory matrix} to map from fine theory models (with $\Delta k = 0.0005\,h\,\text{Mpc}^{-1}$) to the binned and masked data. This additionally accounts for wide-angle effects and radial integral constraints \citep{Beutler:2018vpe,deMattia:2019vdg,Chudaykin:2025aux}. 

We assume in~\cref{eqn:noise} an inverse number density of $\bar{n}^{-1}= 4\,085\,(h^{-1}\,\text{Mpc})^3$ for LOWZ, $\bar{n}^{-1}= 5\,970\,(h^{-1}\,\text{Mpc})^3$ for CMASS, and $\bar{n}^{-1}= 68\,400\,(h^{-1}\,\text{Mpc})^3$ for QSO.
To convert the observed redshifts and angles into comoving wavenumbers, and to compute the covariance matrices, we adopt the fiducial flat $\Lambda$CDM cosmology used in previous BOSS/eBOSS analyses, with $h=0.676$ and $\Omega_\mathrm{m}=0.31$; deviations of the true cosmology from this fiducial choice during parameter inference are accounted for via the AP effect described in~\cref{sec:eft}. We compute covariance matrices in the Gaussian limit, including the mask, integral constraints, wide-angle effects, and varying noise properties, as described in refs.~\cite{Philcox:2024rqr,Chudaykin:2025aux} (shown to be an excellent approximation in ref.~\cite{Wadekar:2020hax}). 

The galaxy power spectrum multipoles are modelled with the one-loop EFTofLSS description of~\cref{sec:eft}, including the IR resummation based on the wiggle/no-wiggle split, as implemented in our modified version of the \texttt{pbj} code~\cite{Byun:2020rgl,Moretti:2023drg}.\footnote{\href{https://gitlab.com/chiaramoretti/pbj}{https://gitlab.com/chiaramoretti/pbj}} Parameter inference is performed with the importance nested sampler \texttt{nautilus}~\cite{Lange:2023ydq}\footnote{
\href{https://github.com/johannesulf/nautilus}{https://github.com/johannesulf/nautilus}} with 6\,000 live points and a target effective sample size $n_\text{eff}=25\,000$.

For the standard cosmological parameters, we vary $\omega_\mathrm{c}\equiv\Omega_\mathrm{c}h^2$, $h$, and $\ln\left(10^{10}A_\mathrm{s}\right)$ with uniform priors, while $\omega_\mathrm{b}\equiv\Omega_\mathrm{b}h^2$ and $n_\mathrm{s}$ are given Gaussian priors centred on the \textit{Planck} results~\cite{Planck:2018vyg,Carrilho:2022mon}, listed in~\cref{tab:priors_eft}. The sum of neutrino masses is fixed to the minimal normal-hierarchy value $\sum m_\nu = 0.06\,\mathrm{eV}$ and is not varied.

Following refs.~\cite{Ivanov:2021kcd,Philcox:2021kcw,Carrilho:2022mon,Chudaykin:2022nru}, for the bias sector, we vary the linear bias $b_1$, the quadratic bias $b_2$ and the quadratic tidal bias $b_{\mathcal{G}_2}$ independently for each redshift bin, with $b_1$ uniform and $b_2, b_{\mathcal{G}_2}$ given Gaussian priors, as listed in~\cref{tab:priors_bias}. The prior means adopted for $b_{\mathcal{G}_2}$ and $b_{\Gamma_3}$ differ between the two tracers because we follow different conventions from the literature for each. For BOSS, we adopt an uninformative $b_{\mathcal{G}_2}=0$ Gaussian prior together with the coevolution relation $b_{\Gamma_3}=(23/42)(b_1-1)$, dynamically evaluated at the sampled $b_1$ in each bin, following refs.~\cite{Philcox:2021kcw,Carrilho:2022mon}. For the eBOSS QSO sample, we instead follow ref.~\cite{Chudaykin:2022nru} and centre both priors on the coevolution-model predictions evaluated at a fixed fiducial $b_1=2.4$, giving $b_{\mathcal{G}_2}=-(2/7)(b_1-1)\simeq-0.4$ and $b_{\Gamma_3}\simeq0.77$. The remaining nuisance parameters entering the bias expansion, counterterms, and stochastic contributions of~\cref{eqn:delta_g_third_2,eqn:ctr_rsd,eqn:noise} ($b_{\Gamma_3}, c_0, c_2, c_{\nabla^4\delta}, \epsilon_0, \epsilon_2$, $\alpha_P, P_\mathrm{shot,2}$) are marginalised over analytically, see~\cref{tab:priors_marg}. $P_\mathrm{shot,2}$ applies only to the eBOSS QSO bin, where it captures an additional quadrupole shot-noise contribution arising from redshift uncertainties~\cite{Chudaykin:2022nru}.

\begin{table}[t]
    \centering
    \begin{minipage}[t]{0.40\linewidth}
        \vspace{0pt}
        \centering
        \begin{tabular} { l c }
        \noalign{\vskip 3pt}\hline\noalign{\vskip 1.5pt}\hline\noalign{\vskip 5pt}
        \multicolumn{1}{c}{\bf Parameter} &
        \multicolumn{1}{c}{\shortstack{\bf Prior range /\\ \bf (mean, $\sigma$)}}\\
        \noalign{\vskip 3pt}\cline{1-2}\noalign{\vskip 3pt}
        $b_1$ & [0,\,4] \\
        $b_2$ & $(0,\,1)$ \\
        $b_{\mathcal{G}_2}$ & $(0,\,1)_{BOSS}$ \\
           & $(-0.4,\,1)_{eBOSS}$ \\
        \hline\noalign{\vskip 1.5pt}\hline
        \end{tabular}
        \caption{Priors on the varied bias parameters, one set per redshift bin.}
        \label{tab:priors_bias}
    \end{minipage}
    \hfill
    \begin{minipage}[t]{0.58\linewidth}
        \vspace{0pt}
        \centering
        \begin{tabular} { l c }
        \noalign{\vskip 3pt}\hline\noalign{\vskip 1.5pt}\hline\noalign{\vskip 5pt}
        \multicolumn{1}{c}{\bf Parameter} &
        \multicolumn{1}{c}{\shortstack{\bf Prior range /\\ \bf (mean, $\sigma$)}}\\
        \noalign{\vskip 3pt}\cline{1-2}\noalign{\vskip 3pt}
        \multicolumn{2}{c}{BOSS DR12 LOWZ--CMASS} \\
        \hline
        $b_{\Gamma_3}$ & $(23/42(b_1 - 1),\,1)$ \\
        $c_0\,[h^{-1}\,\text{Mpc}]^2$ & $(0,\,30)$ \\
        $c_2\,[h^{-1}\,\text{Mpc}]^2$ & $(30,\,30)$ \\
        $c_{\nabla^4\delta}\,[h^{-1}\,\text{Mpc}]^4$ & $(500,\,500)$ \\
        $\alpha_P$ & $(1,\,2)$ \\
        $\epsilon_0\,[h^{-1}\,\text{Mpc}]^2$ & $(0,\,10)$ \\
        $\epsilon_2\,[h^{-1}\,\text{Mpc}]^2$ & $(0,\,10)$ \\
        \noalign{\vskip 3pt}\cline{1-2}\noalign{\vskip 3pt}
        \multicolumn{2}{c}{eBOSS DR16 QSO} \\
        \hline
        $b_{\Gamma_3}$ & $(0.77,\,1)$ \\
        $c_0\,[h^{-1}\,\text{Mpc}]^2$ & $(20,\,20)$ \\
        $c_2\,[h^{-1}\,\text{Mpc}]^2$ & $(30,\,20)$ \\
        $c_{\nabla^4\delta}\,[h^{-1}\,\text{Mpc}]^4$ & $(0,\,200)$ \\
        $\alpha_P$ & $(1,\,1)$ \\
        $\epsilon_2\,[h^{-1}\,\text{Mpc}]^2$ & $(0,\,5)$ \\
        $P_\mathrm{shot,2}\,[h^{-1}\,\text{Mpc}]^3$ & $(-1000,\,1000)$ \\
        \hline\noalign{\vskip 1.5pt}\hline
        \end{tabular}
        \caption{Priors on the analytically marginalised nuisance parameters, one set per redshift bin. All are required to have Gaussian priors and to enter the model linearly, as this is what makes the analytic marginalisation exact.}
        \label{tab:priors_marg}
    \end{minipage}
\end{table}

\subsubsection{Frequency range for the LSS analysis} \label{sec:freq_range}
We define the interval $[k_\mathrm{min}, k_\mathrm{max}]$ as the range effectively probed by the data. A necessary condition for detecting an oscillation is that at least one full period is contained within this range; we adopt the more conservative requirement of at least two full periods. For a linear oscillatory feature of the form $\sin(\omega_\mathrm{lin} k / k_*)$, this leads to a minimum frequency given by $\omega^\mathrm{min}_\mathrm{lin} = 4\pi k_* / (k_\mathrm{max} - k_\mathrm{min})$. In the case of logarithmic oscillations, $\sin(\omega_\mathrm{log} \ln(k / k_*))$, the minimum frequency is instead determined by the ratio between the two extremes of the observable $k$-range, $\omega^\mathrm{min}_\mathrm{log} = 4\pi / \ln(k_\mathrm{max} / k_\mathrm{min})$. The same criterion applies to BUMP and PSC, replacing $\omega_\mathrm{lin}$ and $\omega_\mathrm{log}$ with $\omega_\mathrm{bump}$ and $\omega_\mathrm{psc}$. This is a factor of two more conservative than the single-period criterion used in ref.~\cite{Calderon:2025xod}.

The maximum frequency is set by the sampling resolution of the data, $\Delta k$. For linear oscillations, the Nyquist frequency is $\omega^\mathrm{max}_\mathrm{lin} = \pi k_*/\Delta k$. For logarithmic oscillations, using $\Delta(\ln k)\simeq\Delta k/k$, it becomes $\omega^\mathrm{max}_\mathrm{log} = \pi k/\Delta k$, which is $k$-dependent and most restrictive at small $k$. This argument is based on the data-vector sampling alone and does not include the survey window function, which is expected to further suppress sensitivity to the highest frequencies; see~\cref{app:window}.

These criteria ensure consistency between the theoretical model and the resolution of the data. \textit{BOSS} and \textit{BOSS+eBOSS} span slightly different $k$-ranges (\cref{sec:lss_data}); rather than adopting dataset-specific bounds, we conservatively use a single chosen frequency range for both, evaluating the expressions above on the narrower $k$-range. For LIN and BUMP, this reproduces almost exactly the adopted range $\log_{10}\omega_X\in[0.68,1.97]$. 
For LOG and PSC, the minimum-frequency formula gives $\log_{10}\omega_X=0.46$; dedicated tests with a wider prior show that the amplitude constraints degrade below $0.6$, so we raise the adopted lower bound to $0.6$. The upper bound for LOG and PSC cannot be fixed directly by the Nyquist expression, which is $k$-dependent: our adopted upper bound $\omega_X=10^{2.1}$ equals $\omega^\mathrm{max}_\mathrm{log}$ at $k\simeq0.10\,h\,\mathrm{Mpc}^{-1}$, so at smaller $k$ the highest frequencies in the prior are not resolved by the data binning. We retain this bound since dedicated runs with the upper bound lowered give the same posterior over the common frequency range.

\begin{figure}
    \centering
    \includegraphics[width=\linewidth]{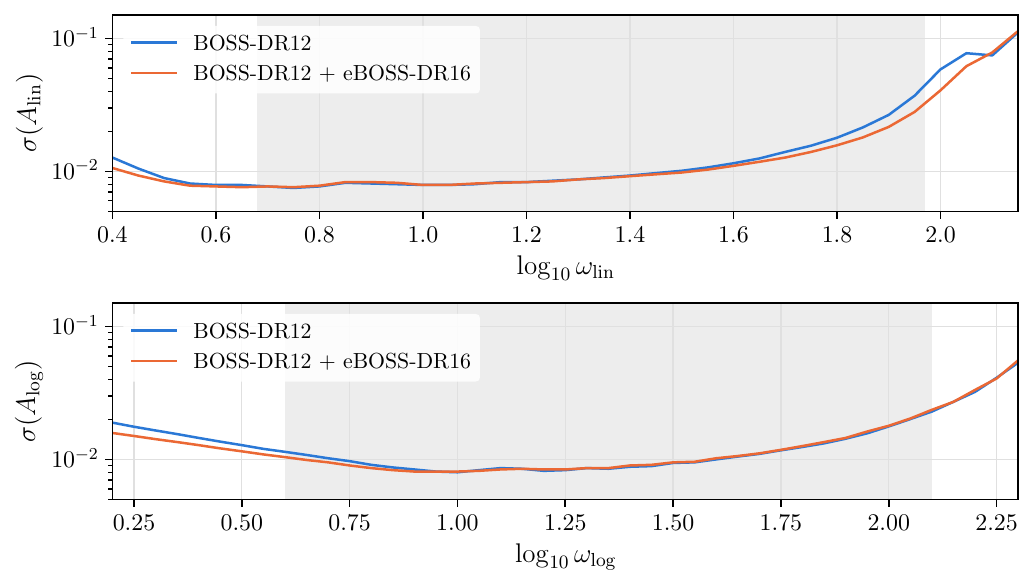}
    \caption{Expected uncertainty on the feature amplitude as a function of frequency for the LIN (left) and LOG (right) templates, for \textit{BOSS} (blue) and \textit{BOSS+eBOSS} (orange), from noise-free mock data at the $\Lambda$CDM best fit analysed with the full likelihood, including the theory matrix, at fixed cosmological and bias parameters. The shaded bands show the adopted frequency priors.}
    \label{fig:sigmaA}
\end{figure}

To check that these ranges reflect the actual sensitivity of the data, including the survey window and binning encoded in the theory matrix, we compute the expected uncertainty on the feature amplitude, $\sigma(A_X)$, as a function of frequency, from noise-free mock data generated at the $\Lambda$CDM best fit and analysed with the full likelihood at fixed cosmological and bias parameters (\cref{fig:sigmaA}). For LOG, $\sigma(A_\mathrm{log})$ varies only mildly over the adopted range, from $\simeq0.008$ at $\log_{10}\omega_\mathrm{log}\simeq1$ to $\simeq0.011$ and $\simeq0.023$ at the lower and upper edges of the prior. For LIN, $\sigma(A_\mathrm{lin})$ is flat down to $\log_{10}\omega_\mathrm{lin}\simeq0.5$, but grows above $\log_{10}\omega_\mathrm{lin}\simeq1.75$ and reaches about five times its minimum at the upper edge of the prior, as a consequence of the window suppression discussed in~\cref{app:window}.

\section{Results} \label{sec:results}
In this section we present results for the four representative models introduced in~\cref{sec:theory}: the sharp feature (LIN), resonant (LOG), bump/dip (BUMP), and classical primordial standard clock (PSC) models. Each model leaves distinct imprints on the PPS, and consequently on the observed LSS and CMB anisotropies. For each model we present results for three dataset combinations: \textit{BOSS} and \textit{BOSS+eBOSS} (\cref{sec:results_lss}), and in combination with CMB data (\cref{sec:results_cmb}). For the CMB, we restrict the results shown here to our baseline \textit{Planck} PR4 \texttt{CamSpec} configuration (\cref{sec:cmb_data}).
We additionally use \textit{eBOSS} alone as an auxiliary cross-check on the frequency and robustness of individual candidates.

\subsection{LSS constraints: BOSS and BOSS+eBOSS} \label{sec:results_lss}
\begin{figure}
    \centering
    \includegraphics[width=\linewidth]{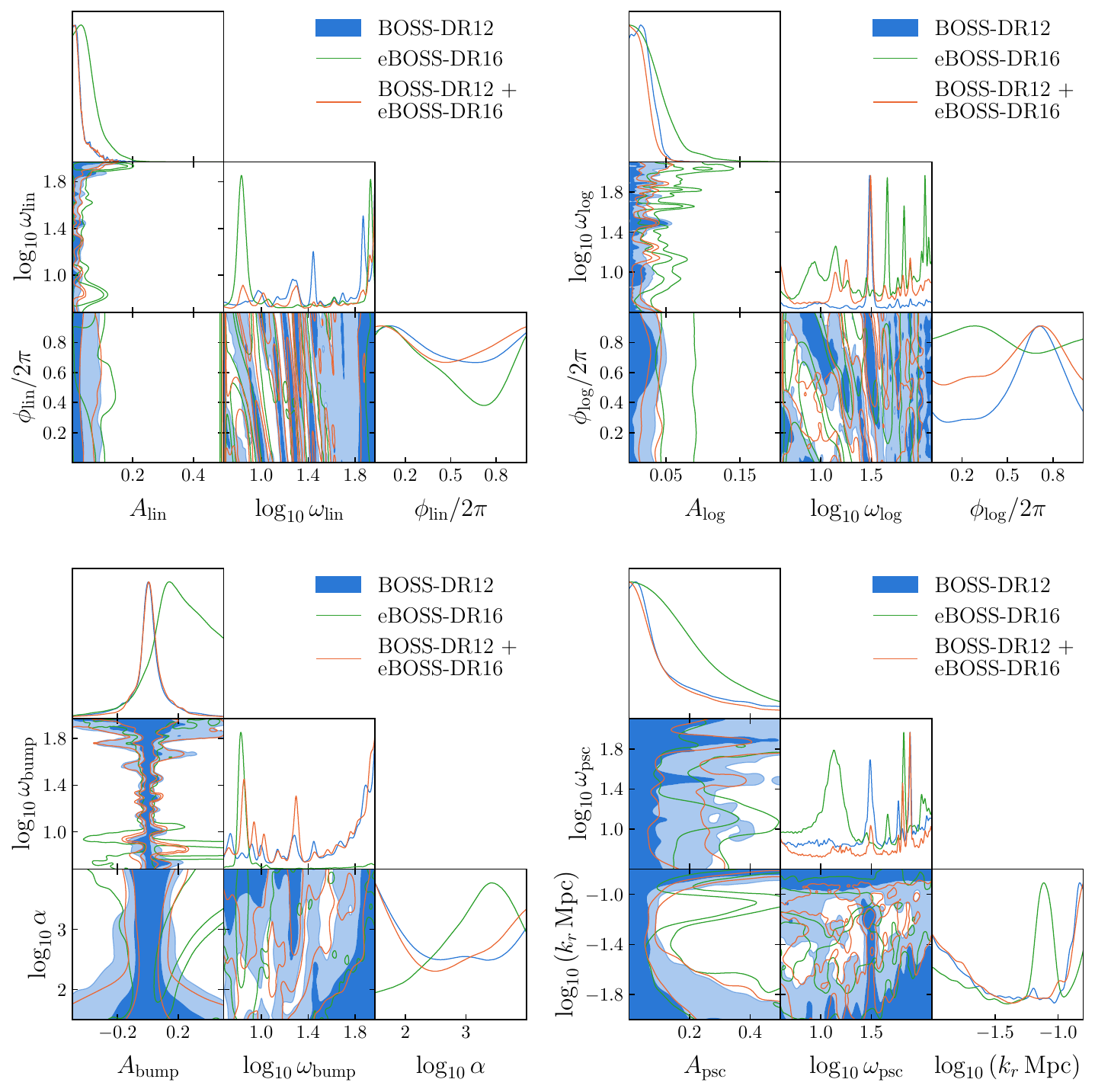}
    \caption{Constraints on the LIN (top left), LOG (top right), BUMP (bottom left), and PSC (bottom right) models. Marginalised joint 68\% and 95\% CL regions of the amplitude, frequency, and phase parameters ($\log_{10}\,\alpha$ for the BUMP model and $\log_{10}\, k_r\,\text{Mpc}$ for the PSC model) using the \textit{BOSS} dataset (blue), \textit{eBOSS} dataset (green), and \textit{BOSS+eBOSS} combination (orange).}
    \label{fig:triangle1}
\end{figure}

We show in~\cref{fig:triangle1} the marginalised posterior distributions at 68\% and 95\% credible level (CL) for the feature parameters of the LIN, LOG, BUMP, and PSC models using the \textit{BOSS}, \textit{eBOSS}, and \textit{BOSS+eBOSS} datasets. The posterior distributions shown here are obtained directly from the nested-sampling chains.
The resulting constraints on the amplitude of each feature template are summarised in~\cref{tab:amplitude}.
\begin{table}[t]
    \centering
    \begin{tabular} {l c c c c}
    \noalign{\vskip 3pt}\hline\noalign{\vskip 1.5pt}\hline\noalign{\vskip 5pt}
    {\bf Dataset} &  {\bf $A_\mathrm{lin}$} &  {\bf $A_\mathrm{log}$} &   {\bf $A_\mathrm{bump}$} &  {\bf $A_\mathrm{psc}$} \\
    \noalign{\vskip 3pt}
    \hline
    \noalign{\vskip 3pt}
    \textit{Planck}                  & $< 0.027$ & $< 0.025$ & $[-0.11,\,0.25]$ & $< 0.29$ \\
    \textit{BOSS}               & $< 0.090$ & $< 0.041$ & $[-0.36,\,0.35]$ & $< 0.42$ \\
    \textit{BOSS + eBOSS}  & $< 0.082$ & $< 0.036$ & $[-0.32,\,0.32]$ & $< 0.40$ \\
    \hline\noalign{\vskip 1.5pt}\hline
    \end{tabular}
    \caption{95\% CL bounds (or 95\% CL interval, where the posterior is not one-sided) on the amplitude of each feature template, for the baseline CMB configuration (from ref.~\cite{Raffaelli:2026bhk}) and the two main LSS datasets.}
    \label{tab:amplitude}
\end{table}
The \textit{Planck} bounds in~\cref{tab:amplitude} are computed over the full frequency range $\log_{10}\omega_X \in [0,2.1]$. For the LSS dataset combinations, the marginalised bound on $A_X$ depends considerably on the frequency prior range: extending the prior to frequencies that are not well constrained by the data inflates the marginalised amplitude bound, since the posterior volume at large $|A_X|$ in that regime is included in the marginalisation. 
The quoted global bounds should therefore be interpreted with caution: they are not representative of the constraining power of the data at every frequency. For LIN, the effect is driven by the high-frequency end of the prior, where the window suppresses the sensitivity of the data (\cref{fig:sigmaA}) and which carries about $37\%$ of the posterior weight: restricting the prior to $\log_{10}\omega_\mathrm{lin}\le1.75$ tightens the 95\% CL bound from $0.090$ and $0.082$ to $A_\mathrm{lin}<0.032$ for both \textit{BOSS} and \textit{BOSS+eBOSS}, stable to within $0.001$ when the cut is moved between $1.70$ and $1.80$.

This also allows a direct comparison with other full-shape analyses that adopt narrower frequency priors. When restricted to the range explored by Chen et al.~\cite{Chen:2026gxp} with $\omega_\mathrm{lin}\in[1.5,40]$ and $\omega_\mathrm{log}\in[2.5,80]$ ($\log_{10}\omega_\mathrm{lin}\in[0.18,1.60]$, $\log_{10}\omega_\mathrm{log}\in[0.40,1.90]$), our bounds tighten to $A_\mathrm{lin}<0.031$--$0.032$ and $A_\mathrm{log}<0.033$--$0.040$ (\textit{BOSS}/\textit{BOSS+eBOSS}), remarkably close to their DESI DR1 constraints of $A_\mathrm{lin}<0.033$ and $A_\mathrm{log}<0.035$, despite the different dataset. When instead restricted to the range of Calderón et al.~\cite{Calderon:2025xod} with $\log_{10}\omega_\mathrm{lin}\in[0.3,1.35]$, $\log_{10}\omega_\mathrm{log}\in[0.7,1.9]$, our bounds are $A_\mathrm{lin}<0.028$--$0.033$, in good agreement with their LSS-only constraint $A_\mathrm{lin}<0.031$ (from the BOSS luminous red galaxy and eBOSS QSO samples, rather than our BOSS LOWZ, CMASS and eBOSS QSO combination); our $A_\mathrm{log}<0.033$--$0.040$ is instead $\sim$1.4--1.6 times weaker than their $A_\mathrm{log}<0.024$. Our more conservative $k_\mathrm{max}=0.20\,h\,\mathrm{Mpc}^{-1}$ for the CMASS sample (versus their $0.23\,h\,\mathrm{Mpc}^{-1}$) reduces the number of sampled oscillation periods by $\sim15\%$ for LIN but only by $\sim3\%$ for LOG, whose period count depends on $\ln(k_\mathrm{max}/k_\mathrm{min})$; it therefore cannot by itself explain why only the LOG bound differs, and the origin of this difference, which may involve the different likelihoods, priors and data selections, remains unclear.

\begin{figure}
    \centering
    \includegraphics[width=\linewidth]{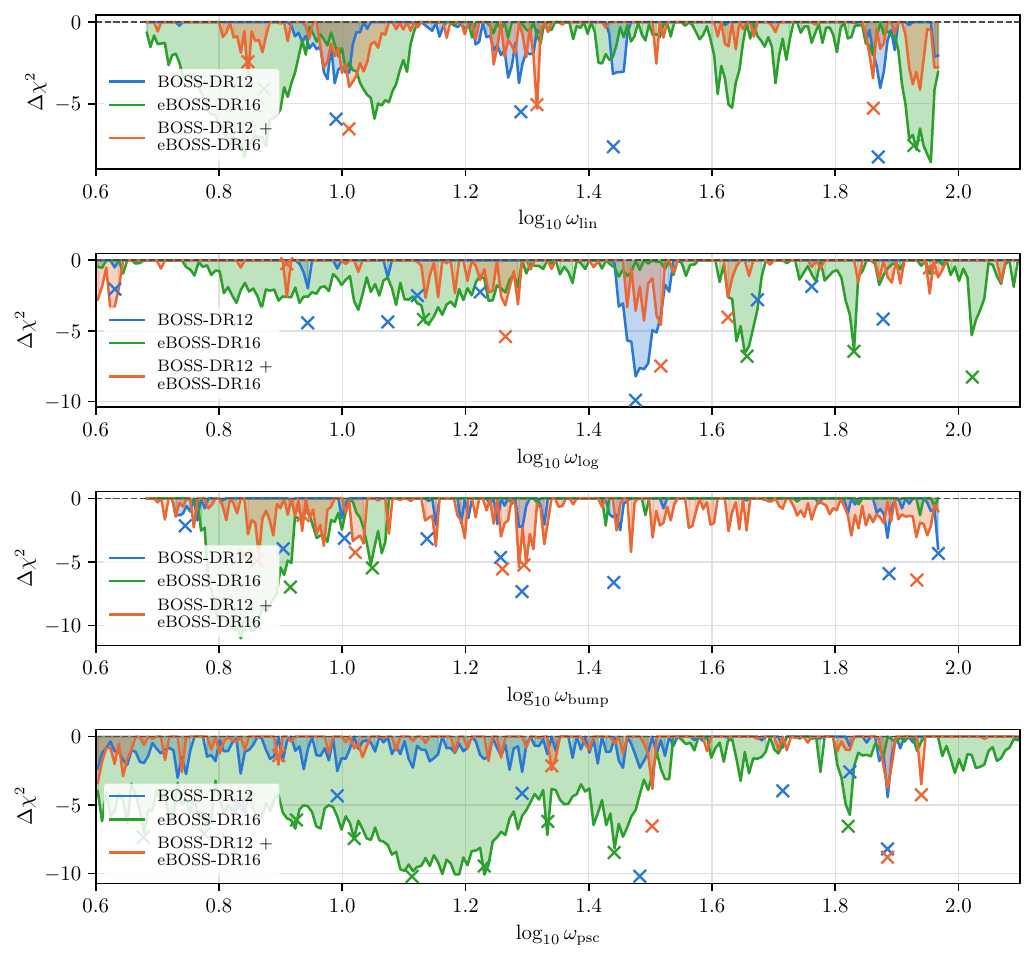}
    \caption{Sample-based $\Delta\chi^2$ reconstruction of the \textit{BOSS} (blue), \textit{eBOSS} (green), and \textit{BOSS+eBOSS} (orange) data in the logarithm of the feature frequency parameter for the LIN, LOG, BUMP, and PSC templates from the top to the bottom, respectively. Markers indicate the frequency and $\Delta\chi^2$ of the dedicated best-fit point found by direct likelihood maximisation in the neighbourhood of each local sample-based optimum.}
    \label{fig:profile}
\end{figure}

\begin{figure}
    \centering
    \includegraphics[width=\linewidth]{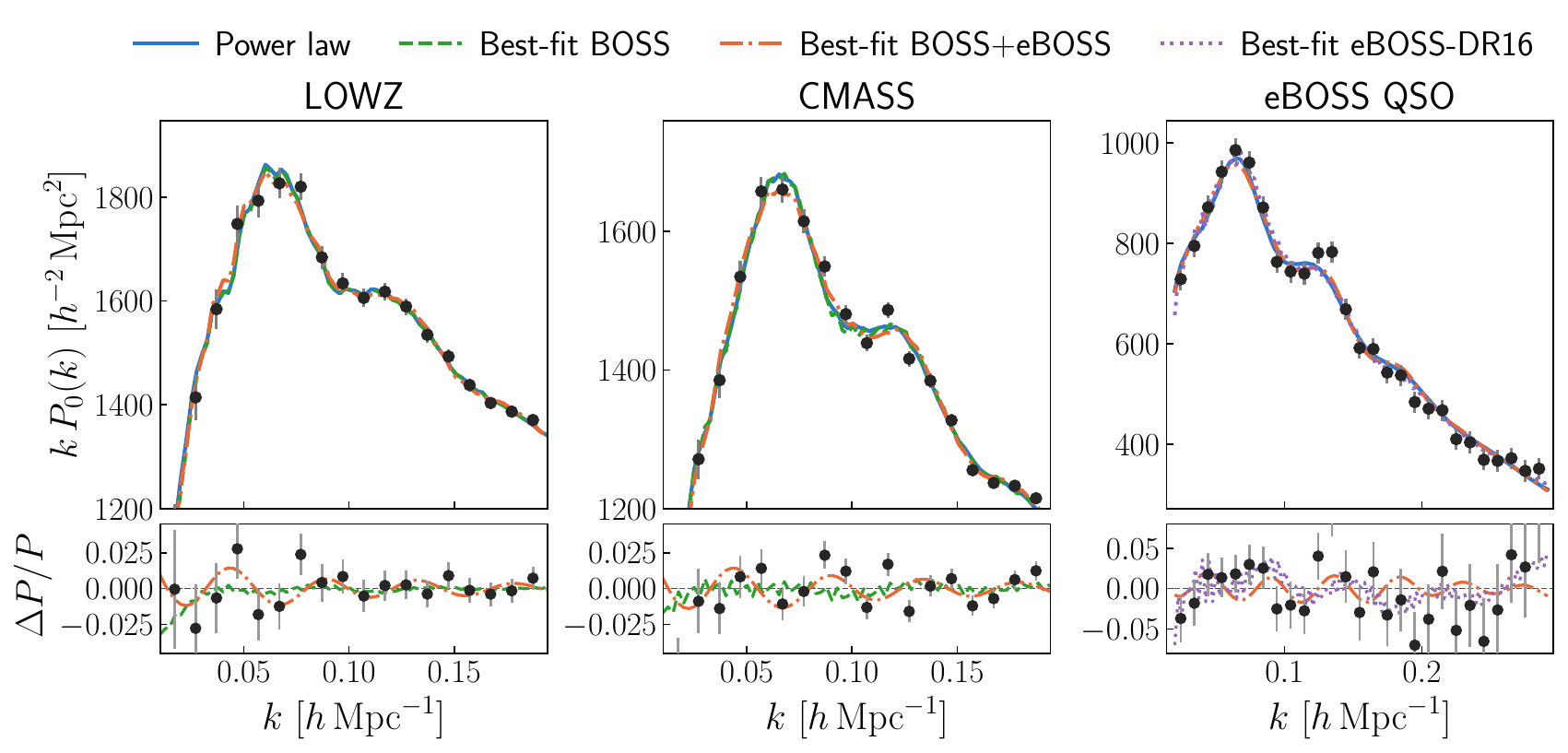}                                                                          \caption{Galaxy power spectrum monopole for LOWZ, CMASS and eBOSS QSO (left to right). Data points (black) are inverse-covariance-weighted averages of the native $\Delta k = 0.0025\,h\,\mathrm{Mpc}^{-1}$ bins into $\Delta k = 0.01\,h\,\mathrm{Mpc}^{-1}$ bins. Theory curves are the window-convolved best fit evaluated at the fine data resolution: the smooth $\Lambda$CDM power-law reference (blue), the LIN best fit to \textit{BOSS} (green), to \textit{BOSS+eBOSS} (orange), and to \textit{eBOSS} (purple). Bottom panels show the relative residual with respect to the power-law reference for both data and best-fit curves.}
    \label{fig:spectra_bf}
\end{figure}

\begin{figure}
    \centering
    \includegraphics[width=\linewidth]{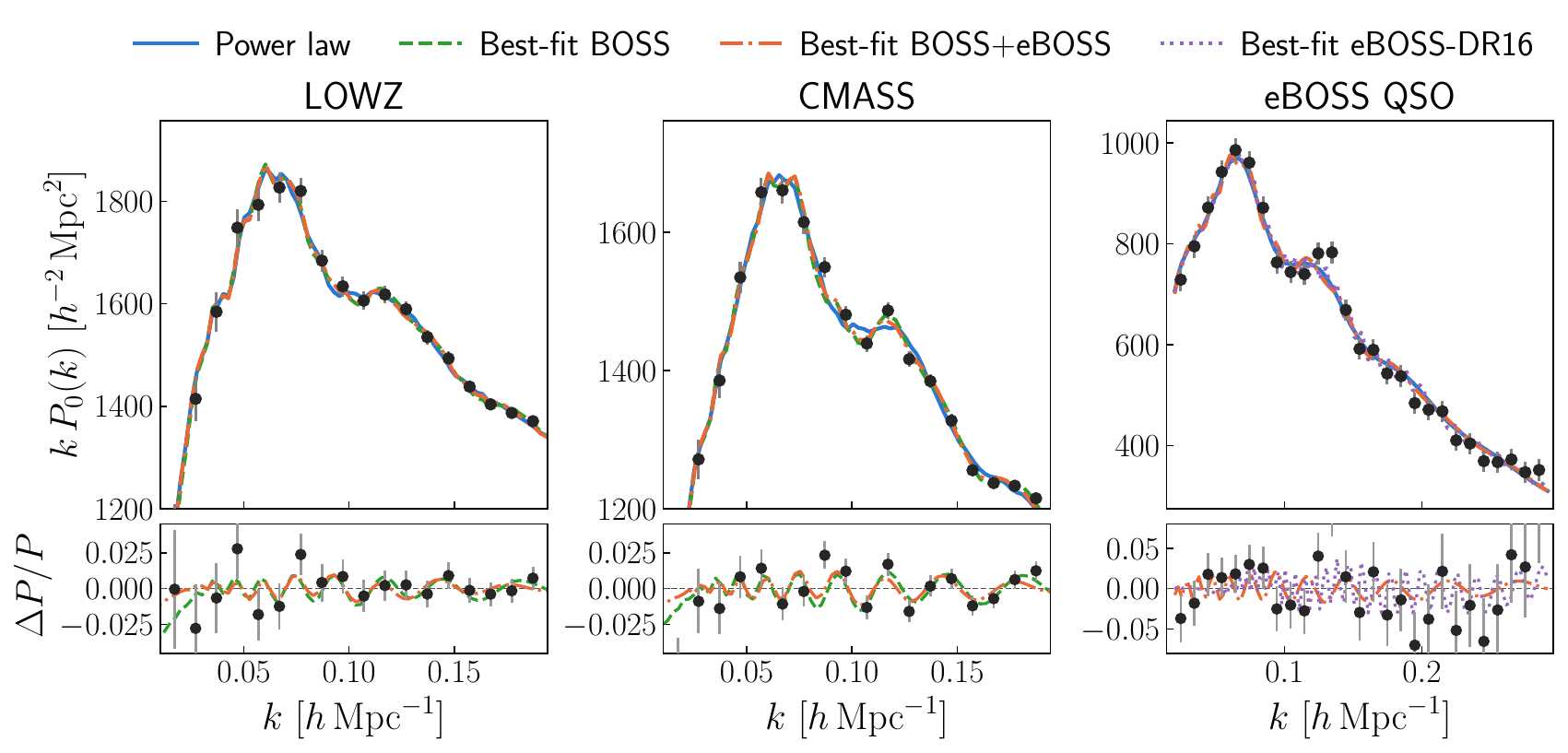}
    \caption{Same as~\cref{fig:spectra_bf} for the LOG model.}
    \label{fig:spectra_bf2}
\end{figure}

\begin{figure}
    \centering
    \includegraphics[width=\linewidth]{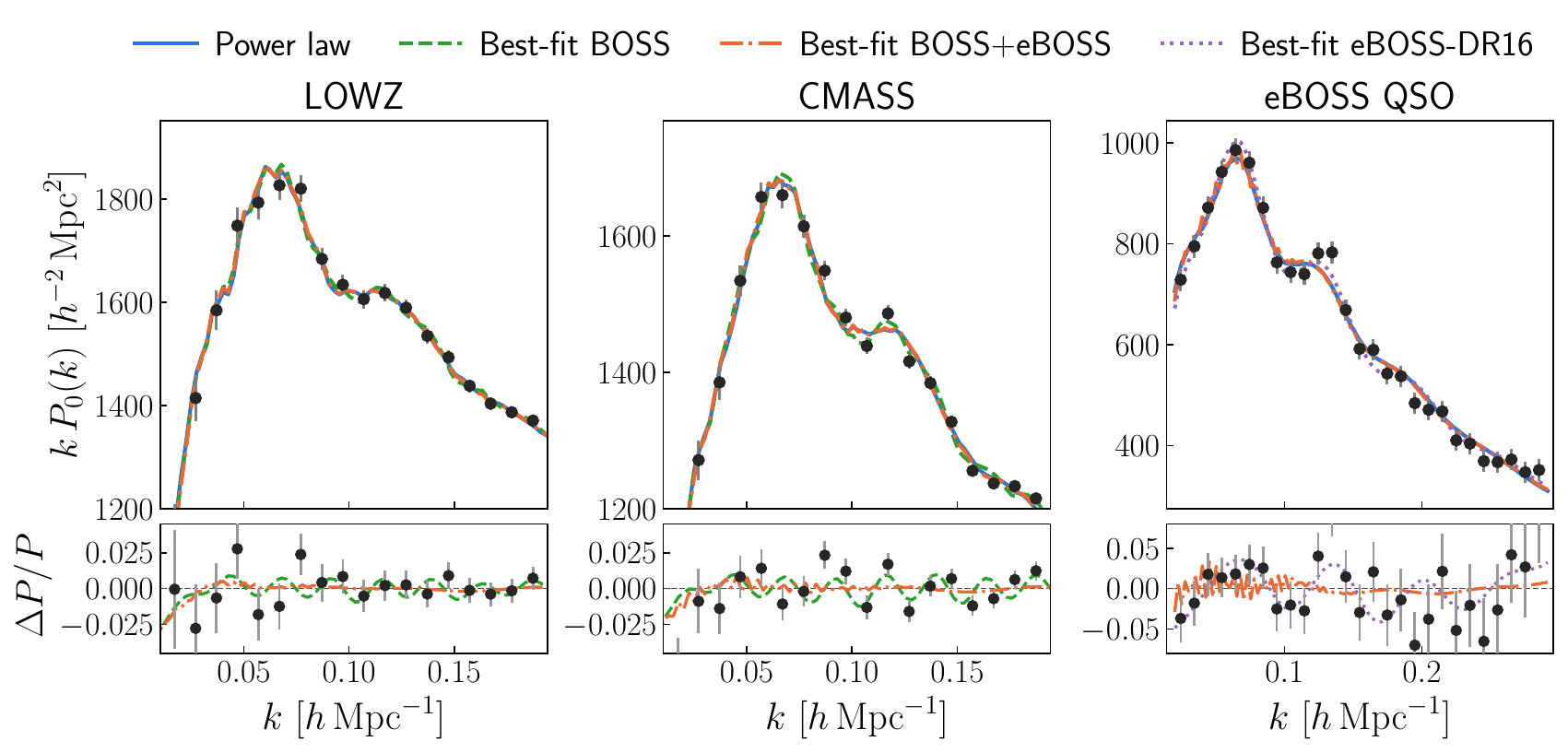}
    \caption{Same as~\cref{fig:spectra_bf} for the BUMP model.}
    \label{fig:spectra_bf3}
\end{figure}

\begin{figure}
    \centering
    \includegraphics[width=\linewidth]{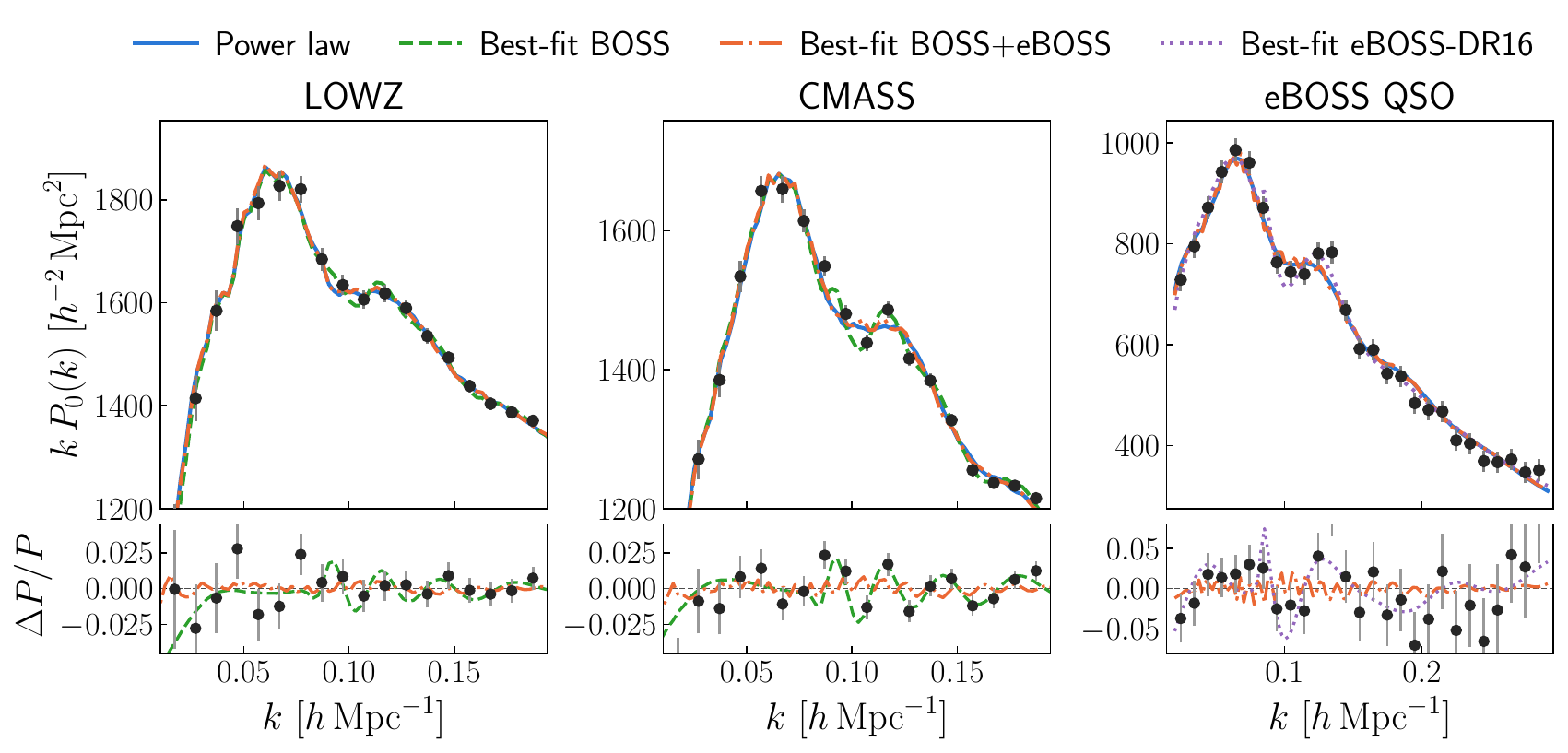}
    \caption{Same as~\cref{fig:spectra_bf} for the PSC model.}
    \label{fig:spectra_bf4}
\end{figure}

In \cref{fig:profile}, we show a sample-based reconstruction of $\Delta\chi^2$ as a function of the logarithm of the feature frequency for the LSS datasets. For each frequency bin, we take the maximum log-likelihood sample already explored by the chain; local maxima identified this way are then refined by a dedicated profile-likelihood maximisation to obtain the best-fit points marked on the curves. This sample-based curve is not a true profile likelihood and can underestimate the depth of narrow local maxima that the chain under-samples, since the sampler is tuned to characterise posterior volume rather than resolve such narrow regions; this explains the occasional offset between the lines and the dedicated best-fit markers in~\cref{fig:profile}.
For a given dataset, the differences for the feature models are calculated with respect to the best-fit likelihood value of the featureless power-law PPS ($\Lambda$CDM) as
\begin{equation}
    \Delta \chi^2 \equiv -2\left(\ln \mathcal{L}\ - \ln \mathcal{L}_\mathrm{\Lambda CDM}^\mathrm{max} \right)
\end{equation}
where values $\Delta \chi^2 < 0$ quantify how much a feature improves the fit to the data. The global maxima for each model and dataset combination are collected in~\cref{tab:maxima} and corresponding window-convolved galaxy power spectrum monopoles are shown in~\cref{fig:spectra_bf,fig:spectra_bf2,fig:spectra_bf3,fig:spectra_bf4}.
Here, $\ln\mathcal{L}$ denotes the analytically marginalised likelihood (i.e.\ the linear nuisance parameters with Gaussian priors are integrated out in closed form, including the resulting Occam term), evaluated at the cosmological and feature parameters that maximise it; this is distinct from a classical profile likelihood, which would minimise $\chi^2$ over the nuisance parameters without that determinant correction.
\begin{table}[t]
    \small
    \centering
    \renewcommand{\arraystretch}{1.4} 
    \setlength{\tabcolsep}{3pt}
    \begin{tabular}{cccccccccc}
    \hline\hline
    \textbf{Model} & \textbf{Dataset} & \boldmath $A_X$ & \boldmath $\log_{10}\omega_X$ & \boldmath $\phi_X/(2\pi)$ & \boldmath $\log_{10}\alpha$ & \boldmath $\log_{10} \left(k_r\,\mathrm{Mpc}\right)$ & \boldmath $\Delta\chi^2$ & \boldmath $\ln B$ & $\bm{S}$ \\
    \hline
    \multirow[c]{3}{*}{LIN}
    & \emph{BOSS}         & $0.066$ & $1.87$ & $0.037$ & -- & -- & $-8.26$ & $-2.6$ & $1.0$ \\[-4pt]
    & \emph{eBOSS}   & $0.14$ & $1.94$ & $0.42$ & -- & -- & $-8.72$ & $-1.1$ & $1.2$ \\[-4pt]
    & \emph{BOSS+eBOSS}   & $0.022$ & $1.02$ & $0.11$ & -- & -- & $-6.54$ & $-2.7$ & $1.1$ \\
    \hline
    \multirow[c]{3}{*}{LOG}
    & \emph{BOSS}         & $0.027$ & $1.48$ & $0.74$ & -- & -- & $-9.91$ & $-2.5$ & $1.6$ \\[-4pt]
    & \emph{eBOSS}   & $0.092$ & $2.03$ & $0.39$ & -- & -- & $-8.27$ & $-1.6$ & $0.6$ \\[-4pt]
    & \emph{BOSS+eBOSS}   & $0.026$ & $1.50$ & $0.80$ & -- & -- & $-7.49$ & $-2.0$ & $0.0$ \\
    \hline
    \multirow[c]{3}{*}{BUMP}
    & \emph{BOSS}         & $0.039$ & $1.29$ & -- & $3.55$ & -- & $-7.34$ & $-1.7$ & $0.8$ \\[-4pt]
    & \emph{eBOSS}   & $0.27$ & $0.85$ & -- & $3.11$ & -- & $-9.96$ & $0.7$ & $1.4$ \\[-4pt]
    & \emph{BOSS+eBOSS}   & $-0.097$ & $1.93$ & -- & $3.29$ & -- & $-6.43$ & $-1.2$ & $0.3$ \\
    \hline
    \multirow[c]{3}{*}{PSC}
    & \emph{BOSS}         & $0.058$ & $1.48$ & $0.80$ & -- & $-1.21$ & $-10.21$ & $-1.1$ & $1.4$ \\[-4pt]
    & \emph{eBOSS}   & $0.20$ & $1.12$ & $0.36$ & -- & $-1.19$ & $-10.23$ & $0.0$ & $1.6$ \\[-4pt]
    & \emph{BOSS+eBOSS}   & $0.40$ & $1.89$ & $0.00$ & -- & $-1.80$ & $-8.82$ & $-1.1$ & $1.0$ \\
    \hline\hline
    \end{tabular}
    \caption{Global maximum frequency and improvement in fit, relative to the featureless power-law PPS, for each model and LSS dataset combination.
    $\ln B\equiv\ln\mathcal{Z}_X-\ln\mathcal{Z}_{\Lambda\mathrm{CDM}}$ is the log Bayes factor against the featureless PPS (negative values favour the featureless model). $S$ is the LEE-corrected global significance of~\cref{eqn:p_global}.}
    \label{tab:maxima}
\end{table}

For each model, multiple local maxima of comparable significance are found across the explored frequency range; \cref{tab:maxima} reports only the global maximum per model and dataset. In the following we discuss all local maxima with $-\Delta\chi^2\geq5$ at the refined best fit, and we consider maxima in different datasets to match when their best-fit frequencies agree within $|\Delta\log_{10}\omega_X|\leq0.03$.
Throughout this discussion we treat two checks as evidence against a dataset-specific fluctuation: consistency of the best-fit frequency across dataset combinations, and an improvement in each of the three \textit{BOSS+eBOSS} redshift bins individually rather than in a single bin.
For \textbf{LIN}, the four \textit{BOSS} maxima at $\log_{10}\omega_\mathrm{lin}\approx 0.98, 1.29, 1.44, 1.87$ reach $-\Delta\chi^2 = 5.5$--$8.3$; two of them recur in \textit{BOSS+eBOSS}, at $1.31$ and $1.87$. The \textit{BOSS+eBOSS} global maximum, at $1.02$ ($-\Delta\chi^2=6.5$), lies $0.04$ above the \textit{BOSS} maximum at $0.98$. The single \textit{eBOSS} maximum, at $1.94$ ($-\Delta\chi^2=8.7$), has no counterpart in either other dataset.
For \textbf{LOG}, only the \textit{BOSS} global maximum exceeds the threshold, at $\log_{10}\omega_\mathrm{log}\approx 1.48$ ($-\Delta\chi^2=9.9$); it recurs at $1.50$ in \textit{BOSS+eBOSS} ($-\Delta\chi^2=7.5$), whose second maximum, at $1.25$ ($-\Delta\chi^2=5.4$), has no \textit{BOSS} counterpart. The three \textit{eBOSS} maxima, at $\log_{10}\omega_\mathrm{log}\approx1.66, 1.82, 2.03$ ($-\Delta\chi^2=6.4$--$8.3$), match neither other dataset.
For \textbf{BUMP}, the three \textit{BOSS} maxima lie at $\log_{10}\omega_\mathrm{bump}\approx 1.29, 1.45, 1.89$ ($-\Delta\chi^2=7.3, 6.6, 5.9$). The \textit{BOSS} global maximum at $1.29$ reappears at $1.30$ in \textit{BOSS+eBOSS} with the same sign of $A_\mathrm{bump}$ (a bump, not a dip) and comparable depth ($-\Delta\chi^2=7.3$ in \textit{BOSS}, $5.2$ in \textit{BOSS+eBOSS}). The deepest \textit{BOSS+eBOSS} maximum, at $\log_{10}\omega_\mathrm{bump}\approx 1.93$ ($-\Delta\chi^2=6.4$), has no \textit{BOSS}-only counterpart and flips sign (a dip). The three \textit{eBOSS} maxima, at $\log_{10}\omega_\mathrm{bump}\approx0.85, 0.90, 1.05$, cluster instead at low frequency.
For \textbf{PSC}, two of the three \textit{BOSS} maxima ($\log_{10}\omega_\mathrm{psc}\approx 0.84, 1.48, 1.88$) recur in \textit{BOSS+eBOSS}, at $1.50$ and $1.89$; the latter is the closest match in the analysis, with $-\Delta\chi^2 = 8.2$ in \textit{BOSS} and $8.8$ in \textit{BOSS+eBOSS}. All nine \textit{eBOSS} maxima, spread over $\log_{10}\omega_\mathrm{psc}\approx0.68$--$1.82$, exceed the threshold ($-\Delta\chi^2=6.1$--$10.2$), but none matches a maximum in the other datasets; the closest, at $1.44$, lies $0.04$ below the \textit{BOSS} maximum at $1.48$. The \textit{eBOSS} global maximum, at $\log_{10}\omega_\mathrm{psc}\approx1.12$ ($-\Delta\chi^2=10.2$), has no counterpart in either other dataset.

At low frequency, $\log_{10}\omega_X\lesssim1$, close to the periodicity of the BAO wiggle, we find maxima for LIN and PSC in \textit{BOSS} ($\log_{10}\omega_X\approx0.98$ and $0.84$) and for BUMP and PSC in \textit{eBOSS} alone.
Fixing the cosmological and bias parameters at their best-fit values and setting the feature amplitude to zero gives a markedly \emph{worse} fit than the featureless $\Lambda$CDM best fit ($\Delta\chi^2=+4.1$ and $+3.1$ for the \textit{BOSS} LIN and PSC maxima): the preferred cosmology at these points fits the data only together with the oscillation. This reflects a degeneracy between the feature and the cosmological parameters that set the BAO wiggle, not independent evidence for a feature.

Three local maxima pass both robustness checks: LIN at $\log_{10}\omega_\mathrm{lin}\approx1.29/1.31$, BUMP at $\log_{10}\omega_\mathrm{bump}\approx1.29/1.30$, and PSC at $\log_{10}\omega_\mathrm{psc}\approx1.88/1.89$; none is reproduced in \textit{eBOSS} alone. LOG's global maxima are frequency-consistent between \textit{BOSS} and \textit{BOSS+eBOSS} ($\approx1.48/1.50$) but, as shown below, dominated by a single redshift bin. No maximum, for any model, is reproduced at matching frequency in all three datasets.

Decomposing the \textit{BOSS+eBOSS} best fit into the contributions of the three redshift bins (LOWZ, CMASS, QSO) shows that each bin improves individually at $\log_{10}\omega_\mathrm{lin}\approx 1.02, 1.31$, $\log_{10}\omega_\mathrm{log}\approx 1.25$, $\log_{10}\omega_\mathrm{bump}\approx 1.30$, and $\log_{10}\omega_\mathrm{psc}\approx 1.89$. At the remaining \textit{BOSS+eBOSS} maxima of LIN ($1.87$), LOG ($1.50$) and PSC ($1.50$) the improvement comes almost entirely from CMASS, with LOWZ and QSO contributing little or disfavouring that frequency. Within \textit{BOSS} alone (LOWZ, CMASS), CMASS, which has the largest effective volume, dominates the improvement at every maximum except LIN at $1.44$ and BUMP at $1.45$, where LOWZ contributes most.
                                                                         
\subsection{Look-elsewhere effect} \label{sec:lee}
The local best-fit frequencies identified in~\cref{sec:results_lss} are selected from an extensive scan over a continuous frequency range, and their nominal local significance therefore overstates the true evidence for a feature once the LEE is accounted for. 

Our method follows the Bayesian trial-factor construction of refs.~\cite{Bayer:2020yss,Bayer:2021lhk}. For each local maximum $i$ of the profile likelihood, the local test statistic 
\begin{equation}
    q_L \equiv 2 \ln \sround{ \frac{ \sup_{\bm{\theta} \in \bm{\Theta}_1 }\mathcal{L}\round{\bm{x} \, | \,\bm{\theta}}}{ \sup_{\bm{\theta} \in \bm{\Theta}_0} \mathcal{L} \round{\bm{x}\, | \, \bm{\theta}}} } \,,
\end{equation} 
is corrected for the number of effective trials $N_\mathrm{trials} = V_{\rm prior}/V_{\rm post}$, 
where $\mathcal{L} \round{\bm{x}\,|\,\bm{\theta}}$ is the analytically marginalised likelihood function, while $\bm{\Theta}_0$ and $\bm{\Theta}_1$ represent the parameter space allowed under the null and the alternative hypotheses, and 
\begin{equation}
    V_{\rm post}=(2\pi)^{M/2}\sqrt{\det\Sigma_i}
\end{equation} 
is the local posterior volume of the non-amplitude feature parameters (frequency and phase for LIN and LOG with $M=2$, frequency and $\alpha$ for BUMP with $M=2$, and frequency, phase and $k_r$ for PSC with $M=3$) and $V_{\rm prior}$ is the corresponding prior volume. The amplitude is excluded from this volume, as it is the parameter being tested. $\Sigma_i$ is the local posterior covariance matrix of the non-amplitude feature parameters around the $i$-th local maximum, estimated empirically directly from the nested-sampling posterior.
The corrected test statistic and the global $p$-value follow~\cite{Bayer:2020yss}
\begin{align}
    q_S &= q_L - 2\ln N_\mathrm{trials} + \ln(2\pi q_L) - 2\ln t \,, \\
    p_\mathrm{global} &= 1-\exp\left(-e^{-q_S/2}\right) \,, \label{eqn:p_global}
\end{align}
with $t=1$ for a one-sided amplitude prior (LIN, LOG, PSC) and $t=2$ for a two-sided one (BUMP). \Cref{eqn:p_global} is the continuous generalisation of the \v{S}id\'ak correction derived in ref.~\cite{Bayer:2020yss}, which remains valid in the non-asymptotic regime. We convert it into a global significance $S$ through the Gaussian tail in the same $t$-sided convention, $p_\mathrm{global}=t\,[1-\Phi(S)]$, with $\Phi$ the standard normal cumulative distribution function, and set $S=0$ when $p_\mathrm{global}\geq t/2$. We evaluate this relation exactly rather than through its asymptotic inversion $S^2\simeq q_S-\ln(2\pi q_S)+2\ln t$~\cite{Bayer:2020yss}, which underestimates $S$ by up to $\sim0.4\,\sigma$ for $S\lesssim2$; at the $\sim2.6\,\sigma$ of the CMB analysis~\cite{Raffaelli:2026bhk} the two agree to $0.04\,\sigma$.

Applying the method to the local maxima identified in~\cref{sec:results_lss}, none reaches a global significance of $2\,\sigma$, and most are consistent with zero once the LEE is accounted for. Eight have $S\approx1\,\sigma$ or above. For \textit{BOSS}: $\log_{10}\omega_{\rm log}\approx1.48$, whose local significance of $3.15\,\sigma$ reduces to $S\approx1.6\,\sigma$; $\log_{10}\omega_{\rm psc}\approx1.48$ ($S\approx1.4\,\sigma$); and $\log_{10}\omega_{\rm lin}\approx1.87$ ($S\approx1.0\,\sigma$). For \textit{BOSS+eBOSS}: $\log_{10}\omega_{\rm lin}\approx1.02$ and $\log_{10}\omega_{\rm psc}\approx1.89$ (both $S\approx1.0\,\sigma$). For \textit{eBOSS} alone: $\log_{10}\omega_{\rm psc}\approx1.12$ ($S\approx1.6\,\sigma$), $\log_{10}\omega_{\rm bump}\approx0.85$ ($S\approx1.4\,\sigma$) and $\log_{10}\omega_{\rm lin}\approx1.94$ ($S\approx1.2\,\sigma$); the PSC and BUMP maxima, however, are close to the BAO scale, so neither should be read as evidence for a genuine feature. The three maxima that pass both robustness checks of~\cref{sec:results_lss} have $S\lesssim1\,\sigma$: LIN at $1.29/1.31$ ($S\approx0.2\,\sigma$ in \textit{BOSS}, $0$ in \textit{BOSS+eBOSS}), BUMP at $1.29/1.30$ ($S\approx0.8\,\sigma$ in \textit{BOSS}), and PSC at $1.88/1.89$ ($S\approx0.6\,\sigma$ and $1.0\,\sigma$).

For several candidates found in both \textit{BOSS} and \textit{BOSS+eBOSS} (e.g.\ LIN at $\log_{10}\omega_{\rm lin}\approx1.87$, BUMP at $\log_{10}\omega_{\rm bump}\approx1.29$, and PSC at $\log_{10}\omega_{\rm psc}\approx1.48$), the local improvement decreases by 30--35\% when \textit{eBOSS} sample is added. A signal shared by the two surveys would be expected to persist or strengthen, so this behaviour is more naturally explained by statistical fluctuations localised to the \textit{BOSS} volume.

We stress that every $S$ quoted above is a global significance \emph{within} its own template's frequency scan; we do not further correct for the trial factor of having tested four distinct template families, which would reduce these significances somewhat further.

\subsection{Combination with CMB data} \label{sec:results_cmb}
\begin{figure}
    \centering
    \includegraphics[width=\linewidth]{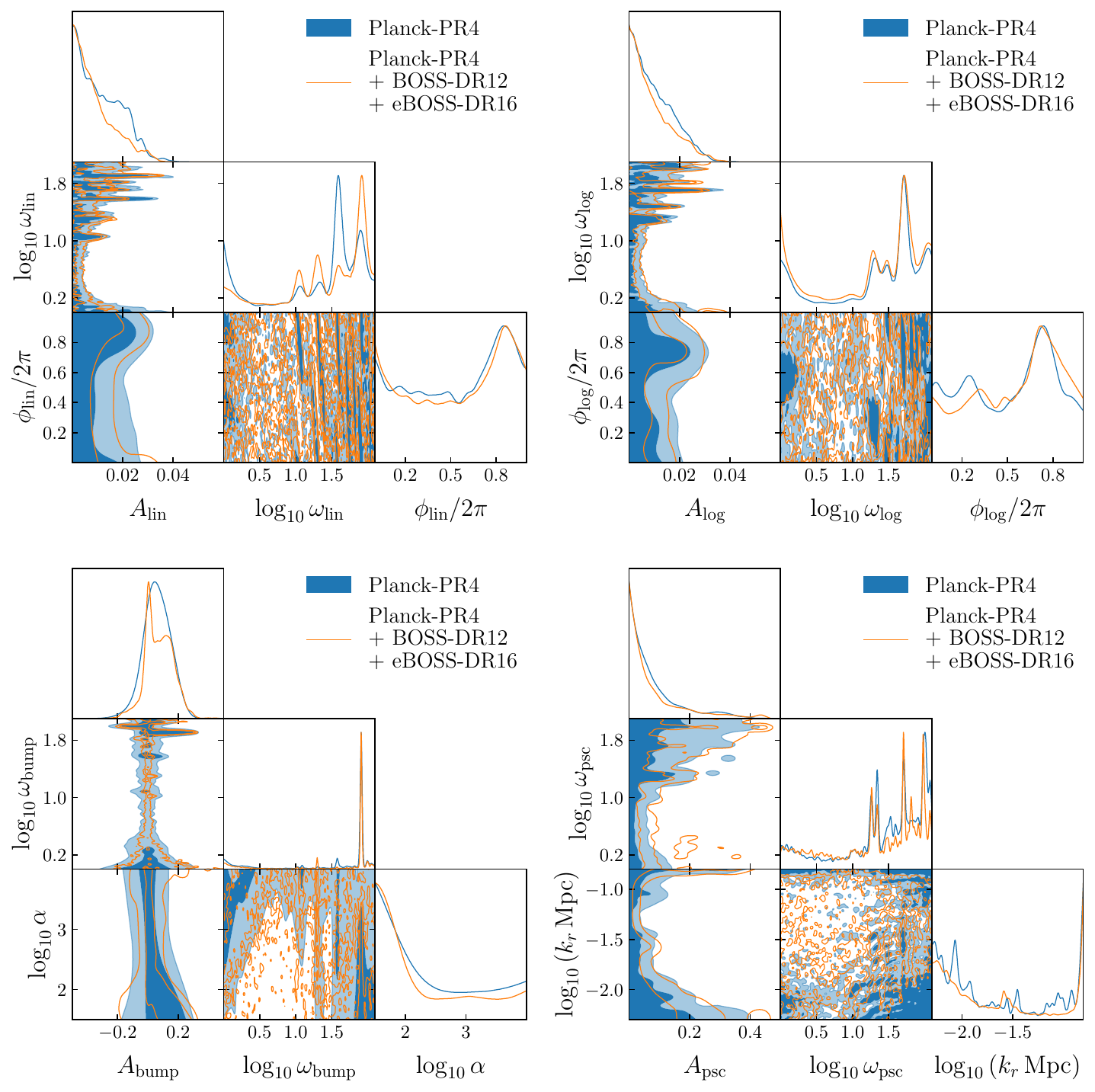}
    \caption{Constraints on the LIN (top left), LOG (top right), BUMP (bottom left), and PSC (bottom right) models. Marginalised joint 68\% and 95\% CL regions of the amplitude, frequency, and phase parameters ($\log_{10}\,\alpha$ for the BUMP model and $\log_{10}\, k_r\,\text{Mpc}$ for the PSC model) using the \textit{Planck} dataset (blue) and \textit{BOSS+eBOSS+Planck} combination (orange).}
    \label{fig:triangle2}
\end{figure}

The marginalised posteriors for \textit{Planck} alone and \textit{BOSS+eBOSS+Planck} are compared in~\cref{fig:triangle2}; as~\cref{tab:amplitude2} shows, combining with the LSS data does not meaningfully tighten the amplitude bounds relative to \textit{Planck} alone. Decomposing the global best fit of the joint chain into its CMB and LSS (\textit{BOSS+eBOSS}) contributions shows that, for all four models, the improvement is almost entirely of CMB origin: LIN's joint best fit, at $\log_{10}\omega_\mathrm{lin}\approx1.91$, has $\Delta\chi^2_\mathrm{CMB}=-8.6$ against $\Delta\chi^2_\mathrm{LSS}=+0.7$ ($\Delta\chi^2_\mathrm{tot}=-7.9$); LOG, at $\log_{10}\omega_\mathrm{log}\approx1.71$, has $-7.7$ against $+1.9$ ($-5.8$ total); BUMP, at $\log_{10}\omega_\mathrm{bump}\approx1.90$, has $-8.5$ against $+1.0$ ($-7.5$ total); PSC, at $\log_{10}\omega_\mathrm{psc}\approx1.70$, has $-12.0$ against $+2.2$ ($-9.8$ total). In each case the LSS data do not themselves favour the jointly preferred frequency.

None of the \textit{BOSS+eBOSS} local maxima of~\cref{sec:results_lss} is reinforced by the combination: the PSC maximum at $\log_{10}\omega_\mathrm{psc}\approx1.89$, the deepest of the three candidates that pass both of our robustness checks, has $\Delta\chi^2_\mathrm{LSS}=+3.0$ in the joint fit, and LIN's deepest \textit{BOSS} maximum ($\log_{10}\omega_\mathrm{lin}\approx1.87$) has $\Delta\chi^2_\mathrm{LSS}=+1.5$. BUMP's deepest \textit{BOSS+eBOSS} maximum ($\log_{10}\omega_\mathrm{bump}\approx1.93$, itself new and unmatched in \textit{BOSS} alone, see~\cref{sec:results_lss}) behaves the same way: $\Delta\chi^2_\mathrm{LSS}=+0.6$ in the joint fit. A partial exception is LOG, whose \textit{BOSS}/\textit{BOSS+eBOSS} maxima ($\log_{10}\omega_\mathrm{log}\approx1.48/1.50$) retain a mildly favourable LSS component ($\Delta\chi^2_\mathrm{LSS}=-2.1$) even though the joint-analysis global best fit sits instead at the deeper, CMB-dominated mode. Overall, the combination provides no evidence for LIN, LOG, BUMP or PSC beyond what \textit{Planck} alone already shows, and does not corroborate any \textit{BOSS+eBOSS} candidate.
Repeating the LEE correction directly on the joint chain at each of these four best-fit frequencies, the significances are further reduced in every case.

Calderón et al.~\cite{Calderon:2025xod} report a substantial tightening of their LIN and LOG bounds when combining with CMB data, unlike the negligible change we find in~\cref{tab:amplitude2}. They use the \textit{Planck} \texttt{Plik} PR3 likelihood, which the companion analysis~\cite{Raffaelli:2026bhk} finds more susceptible to low-frequency features than our baseline \texttt{CamSpec} PR4: its amplitude posterior is not peaked at zero, so combining with LSS data pulls it towards smaller $A_X$, mimicking an improvement that \texttt{CamSpec} PR4 does not show.   
\begin{table}[t]
    \centering
    \begin{tabular} {l c c c c}
    \noalign{\vskip 3pt}\hline\noalign{\vskip 1.5pt}\hline\noalign{\vskip 5pt}
    {\bf Dataset} &  {\bf $A_\mathrm{lin}$} &  {\bf $A_\mathrm{log}$} &   {\bf $A_\mathrm{bump}$} &  {\bf $A_\mathrm{psc}$} \\
    \noalign{\vskip 3pt}
    \hline
    \noalign{\vskip 3pt}
    \textit{Planck}                             & $< 0.027$ & $< 0.025$ & $[-0.11,\,0.25]$ & $< 0.29$ \\
    \textit{BOSS + eBOSS + Planck}     & $< 0.026$ & $< 0.025$ & $[-0.11,\,0.25]$ & $< 0.29$ \\
    \hline\noalign{\vskip 1.5pt}\hline
    \end{tabular}
    \caption{95\% CL bounds (or 95\% CL interval, where the posterior is not one-sided) on the amplitude of each feature template, for the baseline CMB configuration and its combination with the LSS datasets.}
    \label{tab:amplitude2}
\end{table}

\section{Conclusions} \label{sec:conclusions}
We presented a full-shape EFTofLSS analysis of four primordial oscillatory feature templates (LIN, LOG, BUMP, PSC) using BOSS DR12 galaxies, eBOSS DR16 quasars, and their combination, together with the combination with \textit{Planck} PR4 CMB data. We find no statistically significant evidence for any of the four templates in any dataset or combination once the LEE is accounted for, consistent with the CMB-only result of the companion analysis Raffaelli et al.~\cite{Raffaelli:2026bhk} and with the independent analyses of refs.~\cite{Calderon:2025xod,Chen:2026gxp}.

Three local maxima pass both of our robustness checks, frequency consistency between \textit{BOSS} and \textit{BOSS+eBOSS} and an improvement in each \textit{BOSS+eBOSS} redshift bin: LIN at $\log_{10}\omega_\mathrm{lin}\approx1.29/1.31$, BUMP at $\log_{10}\omega_\mathrm{bump}\approx1.29/1.30$, and PSC at $\log_{10}\omega_\mathrm{psc}\approx1.88/1.89$. These checks are not independent, since \textit{BOSS} is contained in \textit{BOSS+eBOSS}, none of the three is reproduced in \textit{eBOSS} alone, and all have LEE-corrected significances of at most $\sim1\,\sigma$. LOG's frequency-consistent maximum at $\log_{10}\omega_\mathrm{log}\approx1.48/1.50$, the largest local improvement in \textit{BOSS} ($S\approx1.6\,\sigma$), is instead driven by a single redshift bin. None constitutes significant evidence for a feature.
A recurring low-frequency region, close to the periodicity of the BAO wiggle, appears as an apparent local maximum for several templates in \textit{BOSS} and \textit{eBOSS} alone, and reflects a degeneracy with the cosmological parameters that set the BAO wiggle.
Combining with \textit{Planck} PR4 does not improve on the CMB-only constraints: the joint best fit for each template reproduces the companion analysis's own \texttt{CamSpec} PR4 local maxima, with the LSS data contributing little or nothing to the preferred frequency, 
and the already-modest CMB-only significances are not enhanced by the combination. Restricted to the frequency ranges adopted by refs.~\cite{Chen:2026gxp,Calderon:2025xod}, our \textit{BOSS+eBOSS} amplitude bounds are comparable with theirs (\cref{sec:results_lss}), despite the different survey and redshift range, although they remain weaker than the CMB ones by factors of $1.4$--$3$.

The present analysis is based on BOSS DR12 and eBOSS DR16; DESI DR1 already offers a larger, more precise galaxy and quasar sample, and ref.~\cite{Chen:2026gxp} has performed the same analysis to the LIN and LOG models. A natural extension of this work is to repeat the dedicated frequency-profile search and LEE correction developed here directly on DESI DR1, and to combine the resulting constraints with \textit{Planck}, to test whether any candidate persists across independent, higher-precision galaxy surveys.

\acknowledgments
NB is supported by the Spanish grants PID2023-147306NB-I00 and CEX2023-001292-S (MCIU/AEI/10.13039/501100011033).  
AR acknowledges support from ASI/INFN grant no. 2021-43-HH.0.
OHEP is supported in part by the National Science Foundation under grant number PHY-2609396. Some of the computing for this project was performed on the Sherlock cluster. We would like to thank Stanford University and Stanford Research Computing for providing computational resources and support that contributed to these research results.

\appendix

\section{Redshift-space IR-resummation kernel} \label{app:sigma_rsd}
Here we detail the derivation of the function $\Sigma^{(s)}_X(k,\mu)$, the redshift-space generalisation of the IR-resummation damping function introduced in~\cref{eqn:S_lin_rsd,eqn:Sigma_s_def} of~\cref{sec:eft}.

We set the coordinate system such that
\(\mathbf{k} = (0,0,k)\), \(\mathbf{q} = (q \sqrt{1-\nu^2},0,q\nu)\), and \(\hat{\mathbf{z}} = (\sqrt{1-\mu^2}\cos\phi, \sqrt{1-\mu^2}\sin\phi, \mu)\). It follows that:
\begin{align}
    &\left(\mathbf{k} \cdot \mathbf{q}\right) = k q \nu = k q \nu \,, \\
    &\left(\mathbf{k} \cdot \hat{\mathbf{z}}\right) = k \mu \,, \\
    &\left(\mathbf{q} \cdot \hat{\mathbf{z}}\right) = q\left(\mu\nu + \sqrt{1-\mu^2}\sqrt{1-\nu^2}\cos\phi\right)\,.
\end{align}
Then, the operators entering~\cref{eqn:S_lin_rsd} are
\begin{align}
    &\frac{\left(\mathbf{k} \cdot \mathbf{q}\right)^2}{q^2} = k^2 \nu^2 \,, \\
    &f \frac{\mathbf{k} \cdot \mathbf{q}}{q} \frac{\hat{\mathbf{z}} \cdot \mathbf{q}}{q} \left(\mathbf{k} \cdot \hat{\mathbf{z}}\right) = f k^2 \mu \nu \left(\mu\nu + \sqrt{1-\mu^2}\sqrt{1-\nu^2}\cos\phi\right) \,, \\
    &f^2 \frac{\left(\hat{\mathbf{z}} \cdot \mathbf{q}\right)^2}{q^2} \left(\mathbf{k} \cdot \hat{\mathbf{z}}\right)^2 = f^2 k^2 \mu^2 \left(\mu\nu + \sqrt{1-\mu^2}\sqrt{1-\nu^2}\cos\phi\right)^2 \,.
\end{align}
Substituting into~\cref{eqn:S_lin_rsd}, we obtain
\begin{align}
    \mathcal{S}^{(s)} P_L^\mathrm{w} \left(k\right) &= k^2 P_L^\mathrm{w} \left(k\right) \int_0^{k_\mathrm{S}} \frac{\dd q}{\left(2 \pi\right)^3} \, P_L^\mathrm{nw}(q) \int_{-1}^{1} \dd \nu \int_0^{2\pi} \dd \phi \, \notag\\
    &\quad \times \left[\left(1+2f\mu^2+f^2\mu^4\right)\nu^2 \right. \notag\\
    &\quad \left. + 2f\mu\nu\sqrt{1-\mu^2}\sqrt{1-\nu^2}\left(1 + f\mu^2\right)\cos\phi \right. \notag\\
    &\quad \left. + f^2\mu^2(1-\mu^2)(1-\nu^2)\cos^2\phi\right]
    \left[1 - \cos \left(\tilde{\omega}_X q \nu\right)\right] \,.
\end{align}
After integrating over the azimuthal angle \(\phi\), the term proportional to $\cos\phi$ cancels, yielding
\begin{align}
    \mathcal{S}^{(s)} P_L^\mathrm{w} \left(k\right) &= k^2 P_L^\mathrm{w} \left(k\right) \int_0^{k_\mathrm{S}} \frac{\dd q}{\left(2 \pi\right)^2} \, P_L^\mathrm{nw}(q) \int_{-1}^{1} \dd \nu \, \left[\left(1+2f\mu^2+f^2\mu^4\right)\nu^2 \right. \notag\\
    &\quad \left. + \frac{1}{2}\left(f^2\mu^2(1-\mu^2)(1-\nu^2)\right)\right]
    \left[1 - \cos \left(\tilde{\omega}_X q \nu\right)\right] \notag\\
    &= k^2 P_L^\mathrm{w} \left(k\right) \int_0^{k_\mathrm{S}} \frac{\dd q}{\left(2 \pi\right)^2} \, P_L^\mathrm{nw}(q) \int_{-1}^{1} \dd \nu \, \notag\\
    &\quad \times \left[\left(1+2f\mu^2-\frac{f^2\mu^2}{2}+\frac{3f^2\mu^4}{2}\right)\nu^2 + \frac{f^2\mu^2}{2}(1-\mu^2)\right]
    \left[1 - \cos \left(\tilde{\omega}_X q \nu\right)\right] \,.
\end{align}
Integrating also over \(\nu\), we obtain
\begin{align} \label{eqn:Sigma_tot}
    \mathcal{S}^{(s)} P_L^\mathrm{w} \left(k\right) &= P_L^\mathrm{w} \left(k\right) \frac{k^2}{6\pi^2} \int_0^{k_\mathrm{S}} \dd q \, P_L^\mathrm{nw}(q) \notag\\
    &\quad \times \left[\left(1+f\mu^2(2+f)\right) \left(1-j_0(\tilde{\omega}_Xq)+2j_2(\tilde{\omega}_Xq)\right) + 3f^2\mu^2(\mu^2-1) j_2(\tilde{\omega}_Xq)\right] \notag\\
    &\equiv k^2 \Sigma_{{\rm tot},\,X}(k,\mu) P_L^\mathrm{w} \left(k\right) \,,
\end{align}
where
\begin{align}
    &\Sigma_{{\rm tot},\,X}(k,\mu) = \left[1+f\mu^2(2+f)\right] \Sigma_X(k) + f^2\mu^2(\mu^2-1) \delta\Sigma_X(k) \,,\\
    &\Sigma_X(k) = \frac{1}{6\pi^2} \int_0^{k_\mathrm{S}} \dd q \, P_L^\mathrm{nw}(q)
\left[1 - j_0\left(\tilde{\omega}_X q\right) + 2 j_2\left(\tilde{\omega}_X q\right)\right] \,,\\
    &\delta\Sigma_X(k) = \frac{1}{2\pi^2} \int_0^{k_\mathrm{S}} \dd q \, P_L^\mathrm{nw}(q) j_2\left(\tilde{\omega}_X q\right) \,.
\end{align}

\subsection{Redshift-space IR-resummation kernels for logarithmic oscillations} \label{app:log_exact}
The reduction of $\mathcal{S}^{(s)}$ to the closed Bessel-function form of~\cref{eqn:Sigma_tot} for the LOG template relies on freezing the local frequency $\tilde\omega_\mathrm{log}(k')=\omega_\mathrm{log}/k'$ at the external wavenumber $k$, which is accurate only in the soft-loop-momentum limit $q\ll k$, as discussed below~\cref{eqn:s_osc}. Beyond this limit, $\mathbf{q}\cdot\nabla_{\mathbf{k}'}$ acting repeatedly on the LOG wiggly piece does not simply reproduce the same local frequency at every order, since each derivative also acts on the $1/k'^n$ prefactor generated by the previous one, so the naive derivative series does not resum into a simple exponential. Ref.~\cite{Beutler:2019ojk} showed, in real space, that this infinite series can nonetheless be evaluated exactly, order by order in the derivative expansion and resummed.

Since the tensor $\mathcal{P}_{ab}\mathcal{P}_{cd}k^ak^cq^bq^d/q^4$ in~\cref{eqn:S_lin_rsd} is a purely geometric prefactor, independent of the functional form of the wiggly power spectrum being resummed, the derivation of ref.~\cite{Beutler:2019ojk} for the action of $[1-\cosh(\mathbf{q}\cdot\nabla_{\mathbf{k}'})]$ on the LOG wiggly piece $P_\mathrm{log}^\mathrm{w}$ carries over unchanged to redshift space; only the tensor prefactor differs from the real-space case. Explicitly, their result reads
\begin{align} \label{eqn:cosh_log_exact}
    &\left[1-\cosh\!\left(\mathbf{q}\cdot\nabla_{\mathbf{k}'}\right)\right] P_X^{\mathrm{w},\mathrm{log}}(k') \Big\rvert_{\mathbf{k}'=\mathbf{k}} \notag\\
    &= \left\{1-\frac{1}{2}\left[\cos\!\left(\omega_\mathrm{log}\ln\!\left(1+\frac{q\nu}{k}\right)\right)+\cos\!\left(\omega_\mathrm{log}\ln\!\left(1-\frac{q\nu}{k}\right)\right)\right]\right\} P_X^{\mathrm{w},\mathrm{log}}(k) \notag\\
    &\quad - \frac{1}{2}\left\{\sin\!\left[\omega_\mathrm{log}\ln\!\left(1+\frac{q\nu}{k}\right)\right]+\sin\!\left[\omega_\mathrm{log}\ln\!\left(1-\frac{q\nu}{k}\right)\right]\right\} \frac{P_L^\mathrm{nw}(k)}{\omega_\mathrm{log}}\frac{\dd\, \delta P_\zeta^\mathrm{log}(k)}{\dd\ln k} \,,
\end{align}
with $\nu \equiv \hat{\mathbf{q}}\cdot\hat{\mathbf{k}}$, which resums the derivative series for $|q\nu/k|<1$, following ref.~\cite{Ballardini:2024dto}. This result rests on further approximations: the shift is taken along $\hat{\mathbf{k}}$, $|\mathbf{k}+\mathbf{q}|\simeq k+q\nu$, which drops a term $q^2(1-\nu^2)/(2k^2)$ in the argument of the logarithm, of the same order as the curvature term that~\cref{eqn:cosh_log_exact} resums. Substituting~\cref{eqn:cosh_log_exact} into~\cref{eqn:S_lin_rsd} and performing the same azimuthal-angle integral as in the derivation of~\cref{eqn:Sigma_tot} yields
\begin{equation} \label{eqn:S_log_rsd_exact}
    \mathcal{S}^{(s)} P_X^{\mathrm{w},\mathrm{log}}(k) = k^2 \Sigma^{(s)}_\mathrm{log}(k,\mu) P_X^{\mathrm{w},\mathrm{log}}(k) - k^2 \hat\Sigma^{(s)}_\mathrm{log}(k,\mu) \frac{P_L^\mathrm{nw}(k)}{\omega_\mathrm{log}}\frac{\dd\, \delta P_\zeta^\mathrm{log}(k)}{\dd\ln k} \,,
\end{equation}
where
\begin{align}
    \Sigma^{(s)}_\mathrm{log}(k,\mu) &\equiv \frac{1}{4\pi^2} \int_0^{k_\mathrm{S}} \dd q\, P_L^\mathrm{nw}(q) \int_{-1}^1 \dd\nu\, \left\{\left(1+f\mu^2\right)^2\nu^2 + \frac12 f^2\mu^2(1-\mu^2)(1-\nu^2)\right\} \notag\\
    &\qquad\qquad \times \left\{1-\frac{1}{2}\left[\cos\!\left(\omega_\mathrm{log}\ln\!\left(1+\frac{q\nu}{k}\right)\right)+\cos\!\left(\omega_\mathrm{log}\ln\!\left(1-\frac{q\nu}{k}\right)\right)\right]\right\} \,, \label{eqn:Sigma_log_rsd}\\
    \hat\Sigma^{(s)}_\mathrm{log}(k,\mu) &\equiv \frac{1}{4\pi^2} \int_0^{k_\mathrm{S}} \dd q\, P_L^\mathrm{nw}(q) \int_{-1}^1 \dd\nu\, \left\{\left(1+f\mu^2\right)^2\nu^2 + \frac12 f^2\mu^2(1-\mu^2)(1-\nu^2)\right\} \notag\\
    &\qquad\qquad \times \frac{1}{2}\left\{\sin\!\left[\omega_\mathrm{log}\ln\!\left(1+\frac{q\nu}{k}\right)\right]+\sin\!\left[\omega_\mathrm{log}\ln\!\left(1-\frac{q\nu}{k}\right)\right]\right\} \,. \label{eqn:Sigma_hat_log_rsd}
\end{align}
Setting $f=0$ in~\cref{eqn:Sigma_log_rsd,eqn:Sigma_hat_log_rsd} reproduces exactly the real-space result of ref.~\cite{Ballardini:2024dto}, which is a useful check on the tensor reduction. \Cref{eqn:S_log_rsd_exact} avoids freezing $\tilde\omega_\mathrm{log}$ at the external wavenumber, within the same approximations, at the cost of leaving the radial and polar integrals over the loop momentum $\mathbf{q}$ unevaluated, since $\ln(1\mp q\nu/k)$ does not admit a further reduction to elementary or Bessel functions.

Only upon expanding $\ln(1\pm q\nu/k)\simeq \pm q\nu/k$ for soft loop momenta $q\ll k$ (the regime dominating the $\mathcal{S}^{(s)}$ integral) do~\cref{eqn:Sigma_log_rsd,eqn:Sigma_hat_log_rsd} reduce to the approximate closed form of~\cref{eqn:Sigma_tot}, as used in the main text, with $\hat\Sigma^{(s)}_\mathrm{log}$ giving the leading correction beyond it. Ref.~\cite{Beutler:2019ojk} found, in real space, that $\hat\Sigma_\mathrm{log}\ll\Sigma_\mathrm{log}$ for the frequencies and wavenumbers of interest, so that the approximate treatment of the main text is in practice an excellent approximation; we expect the same hierarchy to hold for the redshift-space expressions of~\cref{eqn:Sigma_log_rsd,eqn:Sigma_hat_log_rsd}.

These expressions are not used in our numerical results, which adopt the locally constant frequency approximation of~\cref{eqn:s_osc}.

\section{Results for the standard cosmological and bias parameters} \label{app:cosmo_bias}
\begin{figure}[h!]
    \centering
    \includegraphics[width=\linewidth]{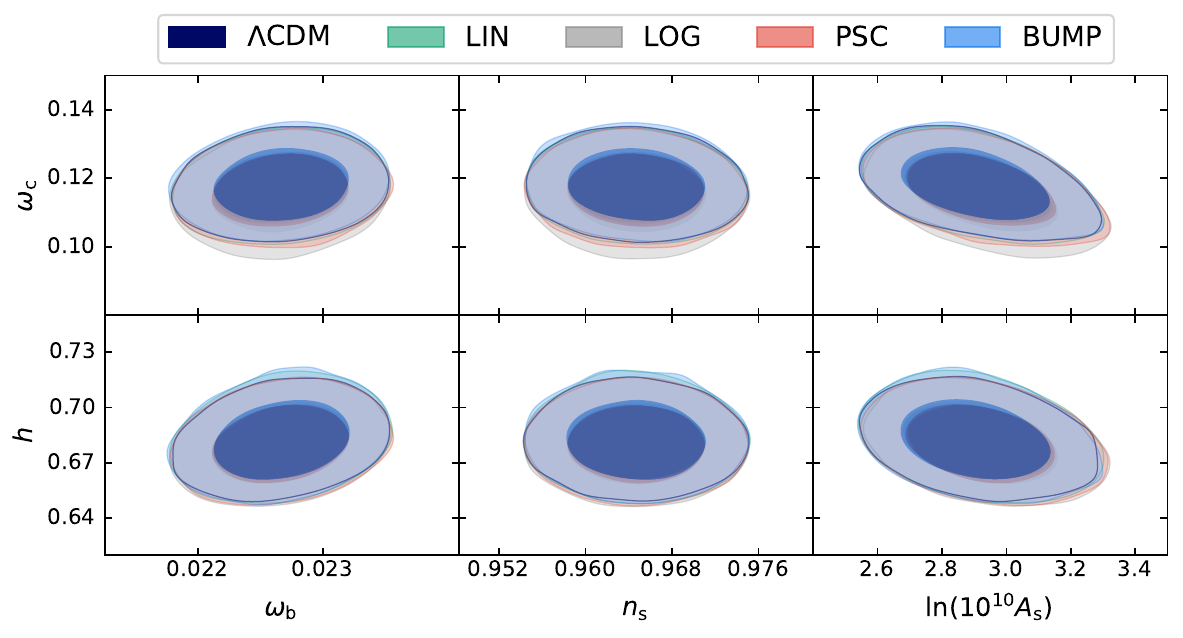}
    \caption{Marginalised joint 68\% and 95\% CL regions in the two-dimensional parameter space of the five standard cosmological parameters using \textit{BOSS}.}
    \label{fig:params} 
\end{figure}

\begin{figure}[h!]
    \centering
    \includegraphics[width=\linewidth]{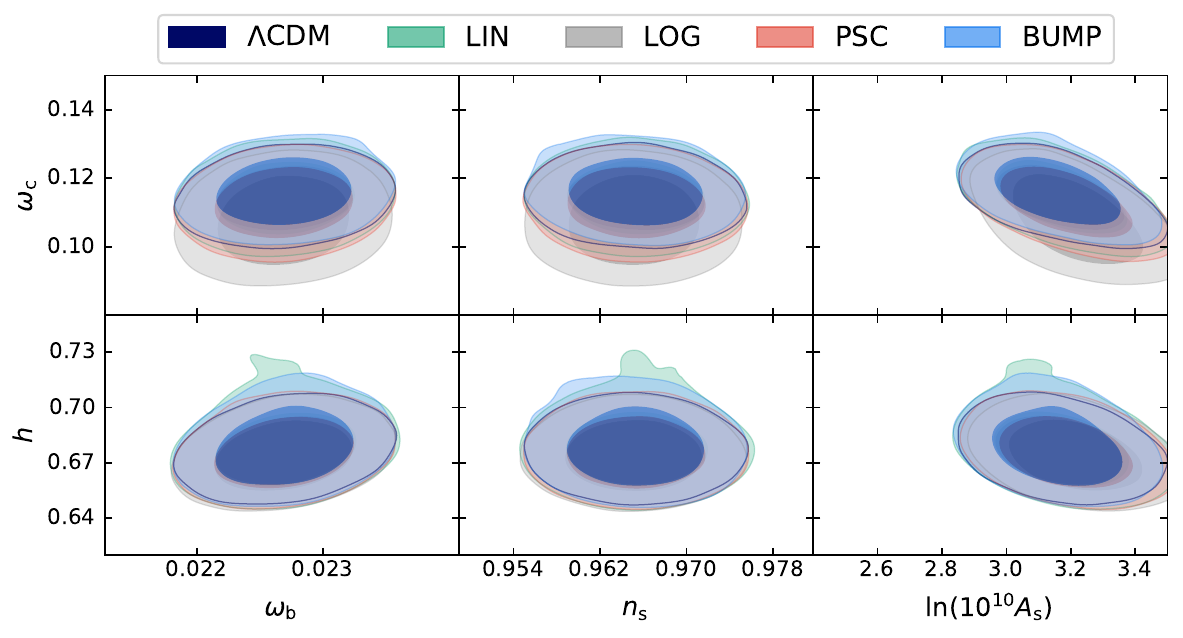}
    \caption{Same as~\cref{fig:params} for \textit{BOSS+eBOSS}.}
    \label{fig:params2} 
\end{figure}

\begin{figure}[ph!]
    \centering
    \includegraphics[width=\linewidth]{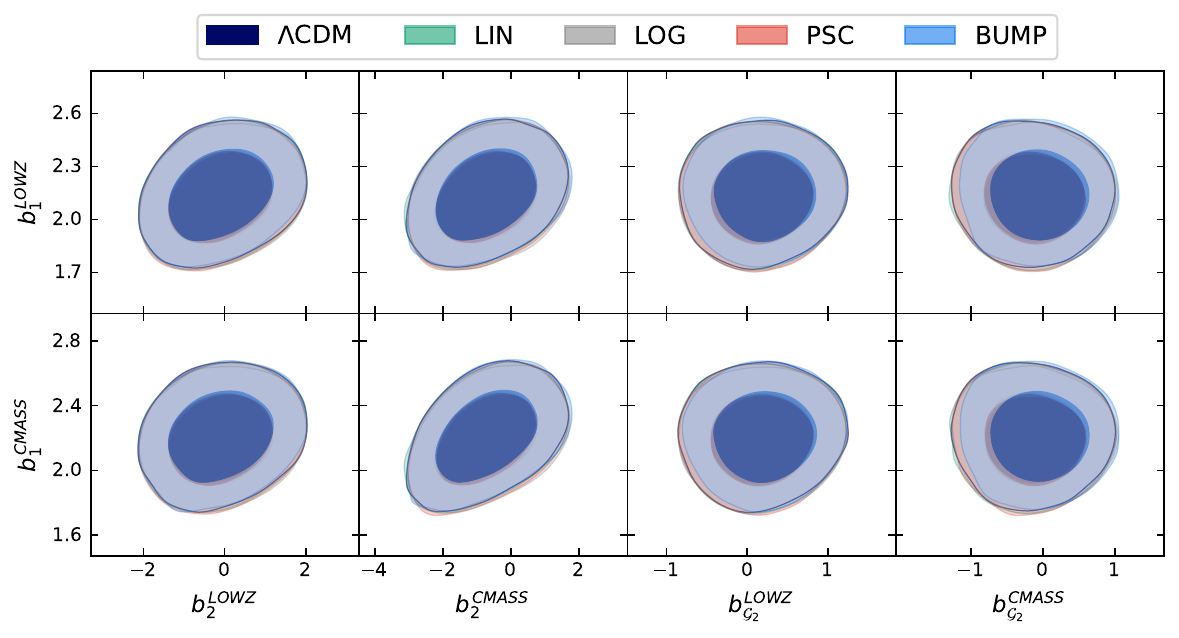}
    \caption{Same as~\cref{fig:params} for the linear bias $b_1$, the quadratic bias $b_2$, and the tidial bias $b_{\mathcal{G}_2}$ at each redshift bin.}
    \label{fig:bias}
\end{figure}

\begin{figure}[ph!]
    \centering
    \includegraphics[width=\linewidth]{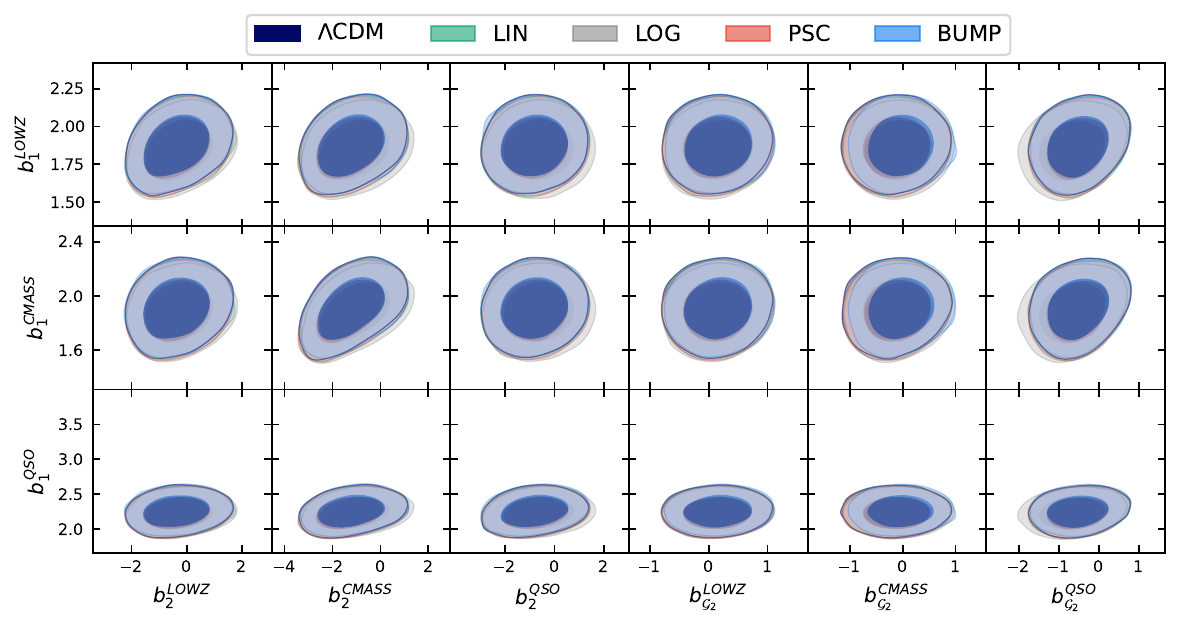}
    \caption{Same as~\cref{fig:bias} for \textit{BOSS+eBOSS}.}
    \label{fig:bias2}
\end{figure}

Here we collect and compare the marginalised posterior distribution of the standard cosmological parameters, that is $\omega_\mathrm{b}$, $\omega_\mathrm{c}$, $h$, $\ln{\left(10^{10}A_\mathrm{s}\right)}$, and $n_\textrm{s}$, in~\cref{fig:params,fig:params2}, and of the bias parameters that are not marginalised analytically, namely $b_1$, $b_2$, and $b_{\mathcal{G}_2}$, in~\cref{fig:bias,fig:bias2}, for \textit{BOSS} and \textit{BOSS+eBOSS}, respectively.
We see that the posterior distributions under the different models for the PPS considered are very stable compared to the results obtained with a power-law PPS within the spatially flat $\Lambda$CDM model.


The constraints on the standard cosmological parameters  for the $\Lambda$CDM model with power-law PPS are reported in~\cref{tab:param}.
\begin{table}[t]    
    \centering
    \footnotesize
    \begin{tabular} { l  c c c c}
    \noalign{\vskip 3pt}\hline\noalign{\vskip 1.5pt}\hline\noalign{\vskip 5pt}
    \multicolumn{1}{c}{\bf Parameter}      & \multicolumn{1}{c}{\textit{BOSS}} & \multicolumn{1}{c}{\textit{BOSS+eBOSS}} & \multicolumn{1}{c}{\textit{BOSS+eBOSS+Planck}} & \multicolumn{1}{c}{\textit{Planck}}\\
    \noalign{\vskip 3pt}\cline{1-5}\noalign{\vskip 3pt}
    $\omega_\mathrm{c}$                 & $0.1176^{+0.0065}_{-0.0072}$      & $0.1143^{+0.0058}_{-0.0066}$            & $0.1195 \pm 0.0009$ & $0.1195\pm0.0010$ \\
    $\omega_\mathrm{b}$                 &  $0.02267 \pm 0.00036$            & $0.02270 \pm 0.00035$         & $0.02221 \pm 0.00013$ & $0.02220\pm0.00013$\\
    $\ln\left(10^{10}A_\mathrm{s}\right)$  & $2.91 \pm 0.15$                   & $3.17 \pm 0.13$                         & $3.046^{+0.011}_{-0.012}$ & $3.046^{+0.011}_{-0.012}$\\
    $n_\mathrm{s}$                         & $0.9647 \pm 0.0042$               & $0.9653 \pm 0.0041$            & $0.9642 \pm 0.0039$ & $0.9644\pm 0.0039$ \\
    $h$                                    & $0.682^{+0.013}_{-0.014}$         & $0.677 \pm 0.012$                      & $0.6733 \pm 0.0040$ & $0.6735\pm 0.0045$ \\
    \hline\noalign{\vskip 1.5pt}\hline
    \end{tabular}
    \caption{Mean values and marginalised constraints at 68\% CL on the standard cosmological parameters for the different datasets for the $\Lambda$CDM model.}
    \label{tab:param}
\end{table}

\section{Window-convolution damping and dephasing of oscillatory features} \label{app:window}
\begin{figure}
    \centering
    \includegraphics[width=\linewidth]{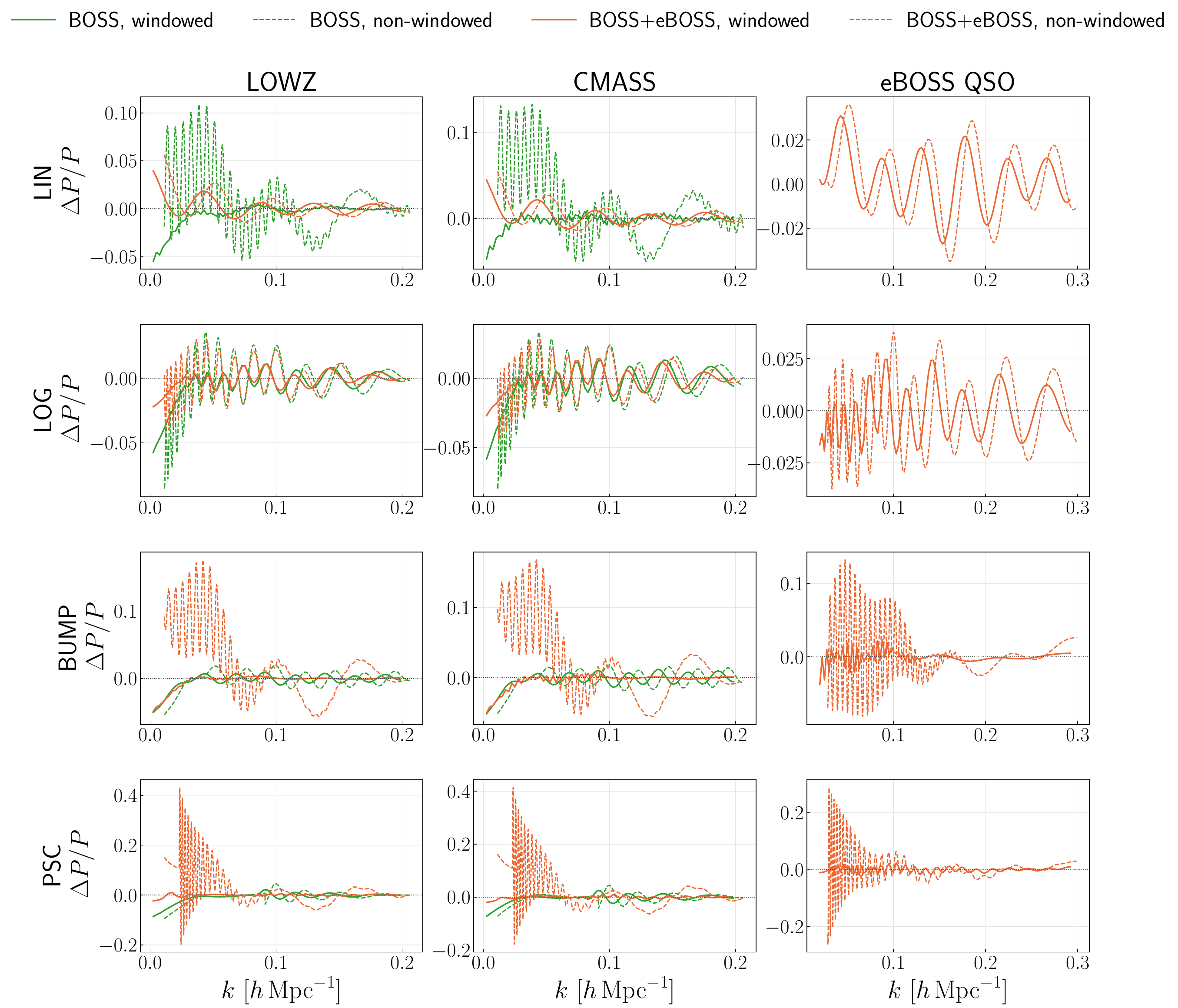}
    \caption{Impact of survey-window convolution on the best-fit curves of~\cref{fig:spectra_bf,fig:spectra_bf2,fig:spectra_bf3,fig:spectra_bf4} on the galaxy power spectrum monopole for LOWZ, CMASS and eBOSS QSO (left to right). For each best fit, the relative deviation from the featureless power-law PPS, $\Delta P/P$, is shown both after window convolution (solid) and without it (dashed).}
    \label{fig:window}
\end{figure}

\Cref{fig:window} compares, for the four templates, the relative deviation from the smooth power-law reference, $\Delta P/P$, computed with (solid) and without (dashed) convolution by the survey window, for the \textit{BOSS} and \textit{BOSS+eBOSS} best fits. Two distinct effects are visible. First, the oscillation amplitude is suppressed by the window, and increasingly so at higher frequency. For the LIN model, for the \textit{BOSS+eBOSS} best fit ($\log_{10}\omega_\mathrm{lin} \approx 1.02$) the windowed and unwindowed curves nearly coincide, while for the higher-frequency \textit{BOSS} best fit ($\log_{10}\omega_\mathrm{lin} \approx 1.87$) the window suppresses the oscillation almost entirely. Second, and less obviously, the windowed and unwindowed curves are visibly out of phase.

Both effects are a generic consequence of convolving with a window kernel that is not symmetric. For an oscillatory mode $u(k')=e^{i\kappa k'}$ convolved with a real kernel $W$,
\begin{equation} \label{eqn:window_1d}
    u_W(k) = \int \mathrm{d}k'\,W(k-k')\,u(k') = \hat W(\kappa)\,e^{i\kappa k} \,, \qquad
    \hat W(\kappa) = \int \mathrm{d}x\,W(x)\,e^{-i\kappa x} \,,
\end{equation}
so the output is rescaled by $|\hat W(\kappa)|$ and shifted in phase by $\arg\hat W(\kappa)$. Writing $W=W_{\rm even}+W_{\rm odd}$, only $W_{\rm even}$ contributes to the real part of $\hat W$ and only $W_{\rm odd}$ to its imaginary part, so a phase shift other than $0$ or $\pi$ requires $W_{\rm odd}\neq0$. \Cref{eqn:window_1d} is only a one-dimensional illustration: the actual multipole window is a mixing matrix $W(k,k')$ that is not translation invariant, and the spherical measure $k'^2\dd k'$ makes the effective kernel asymmetric about $k$ even for an isotropic window, so both suppression and dephasing are generic for any realistic window convolution.
Note that adding a $P_2$ leakage term, in or out of phase with $P_0$, changes the induced shift negligibly, showing that the multipole mixing is subdominant.

\section{Validation with noise-free mock data} \label{app:asimov}
We validate the full analysis pipeline with noise-free mock data. For each LSS dataset, the mock data vector is the $\Lambda$CDM model evaluated at the best fit to the data, with the analytically marginalised nuisance parameters set to their conditional best-fit values; it is analysed with the same covariance, window, theory matrix, priors and sampler settings as the data. By construction, the true feature amplitude is zero and there is no noise.

We repeat the full LIN and LOG analyses on the mock data. No spurious feature is recovered: the maximum-likelihood sample does not improve on the input model, and the posterior on the frequency is smooth. For LIN it increases towards the upper edge of the prior, by a factor of $\sim2$--$3$ with respect to a uniform distribution, as in the data: this reflects the loss of sensitivity shown in~\cref{fig:sigmaA}, not a preference for high frequencies. A mild enhancement at $\log_{10}\omega_\mathrm{lin}\simeq0.85$, close to the BAO frequency, reflects the partial degeneracy with the cosmological parameters that set the BAO feature. The expected 95\% CL bounds are listed in~\cref{tab:asimov}; for LIN restricted to $\log_{10}\omega_\mathrm{lin}\le1.75$, the expected bound, $0.020$, agrees with $\simeq2.45\,\sigma(A_\mathrm{lin})$, the 95\% CL upper limit for an amplitude with free phase. The bounds from the data are a factor of $\sim1.5$--$3$ weaker than the expected ones, consistent with the local maxima discussed in~\cref{sec:results_lss}.
\begin{table}[t]
    \centering
    \begin{tabular}{l c c c c}
    \noalign{\vskip 3pt}\hline\noalign{\vskip 1.5pt}\hline\noalign{\vskip 5pt}
     & \multicolumn{2}{c}{\textit{BOSS}} & \multicolumn{2}{c}{\textit{BOSS+eBOSS}} \\
     & mock & data & mock & data \\
    \noalign{\vskip 3pt}\hline\noalign{\vskip 3pt}
    $A_\mathrm{lin}$ & $<0.039$ & $<0.090$ & $<0.029$ & $<0.082$ \\
    $A_\mathrm{lin}$, $\log_{10}\omega_\mathrm{lin}\le1.75$ & $<0.020$ & $<0.032$ & $<0.019$ & $<0.032$ \\
    $A_\mathrm{log}$ & $<0.024$ & $<0.041$ & $<0.024$ & $<0.036$ \\
    \hline\noalign{\vskip 1.5pt}\hline
    \end{tabular}
    \caption{Expected 95\% CL bounds on the feature amplitude from the noise-free mock data, compared with those from the data (\cref{tab:amplitude}).}
    \label{tab:asimov}
\end{table}

\section{Validation of one-loop integrals with primordial features} \label{app:integrals}
In \texttt{pbj} the one-loop integrals of~\cref{sec:eft} are evaluated with the \texttt{FAST-PT} algorithm~\cite{McEwen:2016fjn,Fang:2016wcf}. Following ref.~\cite{Oddo:2020phd}, the redshift-space one-loop galaxy power spectrum is decomposed into $k$-dependent integrals multiplied by monomials of the bias parameters, the growth rate and $\mu$,
\begin{equation} \label{eqn:loop_decomp}
    P_{g,22}^{(s)\,\text{1-loop}}(k,\mu) = \sum_{X} c_X\, P_X(k) \,, \qquad
    P_{g,13}^{(s)\,\text{1-loop}}(k,\mu) = Z_1(\mu) \sum_{Y} d_Y\, P_{Z_1 Y}(k) \,,
\end{equation}
where the 28 coefficients $c_X \in \{b_1^2,\allowbreak\, b_1 b_2,\allowbreak\, b_1 b_{\mathcal{G}_2},\allowbreak\, \dots,\allowbreak\, \mu^6 f^3 b_1,\allowbreak\, \mu^8 f^4\}$ and the seven $d_Y \in \{b_1,\allowbreak\, b_{\Gamma_3},\allowbreak\, b_{\mathcal{G}_2},\allowbreak\, \mu^2 f,\allowbreak\, \mu^2 f b_1,\allowbreak\, \mu^2 f^2,\allowbreak\, \mu^4 f^2\}$ also label the corresponding integrals (e.g.\ $P_{\mu^2 f b_1}$ multiplies $\mu^2 f b_1$). In the following, $P_{22}$ and $P_{13}$ refer to these two sets of integrals. Five of the $P_X$ and two of the $P_{Z_1 Y}$ are exact multiples of other terms, leaving 23 and 5 independent integrals, respectively. \texttt{FAST-PT} writes every $P_X$ as a linear combination of seven tensor integrals
\begin{equation}
    J_{\alpha\beta\ell}(k) = \int \frac{\dd^3 q}{(2\pi)^{3}} \, q^\alpha p^\beta\, \mathcal{L}_\ell(\hat{\mathbf{q}}\cdot\hat{\mathbf{p}})\, P_L(q) P_L(p) \,,
\end{equation}
with $\mathbf{p} = \mathbf{k}-\mathbf{q}$, each computed with FFTLog~\cite{Hamilton:1999uv} on a logarithmically spaced table of $P_L$ that is extended with power laws and zero-padded; the $P_{Z_1 Y}$ are instead one-dimensional convolutions in $\ln q$, evaluated with fast Fourier transforms. These routines have been validated for featureless $\Lambda$CDM spectra with BAO, but an oscillating input is a more demanding case: the integrand of $P_{22}$ oscillates in both $q$ and $p$, and the finite sampling of the input table, the zero padding and the low-pass filter applied to the Fourier coefficients can alias or suppress the oscillations and spoil the IR cancellations between the tensors.

We therefore compare \texttt{FAST-PT} with a direct two-dimensional integration of the quantity that enters the IR-resummed NLO galaxy power spectrum of~\cref{sec:eft}, namely the feature part of each integral,
\begin{equation} \label{eqn:loop_feature_part}
    \Delta P_X(k) \equiv P_X\!\left[P_L^\mathrm{nw}\left(1+\delta P_\zeta^X\right)\right] - P_X\!\left[P_L^\mathrm{nw}\right] \,,
\end{equation}
and analogously for $\Delta P_{Z_1 Y}$; the BAO wiggles are omitted, consistently with neglecting the BAO--feature cross term in~\cref{eqn:wiggle_spectrum}.

\subsection{Direct integration settings}
The reference kernels are derived independently of \texttt{pbj}, from the SPT recursion relations for $F_n$ and $G_n$, the redshift-space kernels $Z_n$, and the bias operators of~\cref{eqn:delta_g_third_2}, averaged analytically over the azimuth of $\mathbf{q}$ around $\mathbf{k}$, with the same IR regularisation \texttt{pbj} inherits unchanged from \texttt{FAST-PT}'s treatment of the matter integrals~\cite{McEwen:2016fjn}: removal of the $P_{13}$ displacement term $-k^2\sigma_v^2\left[1+f(2+f)\mu^2\right] Z_1^2(\mu) P_L(k)$, with
\begin{equation}
    \sigma_v^2 = \frac{1}{6\pi^2} \int \dd q\, P_L(q) \,,
\end{equation}
and subtraction from $P_{22}$ of its IR-divergent piece $\propto k^2\sigma_v^2 P_L(k)$, together with the constant $k\to0$ limit of $P_{b_2 b_2}$. Both sides integrate over the same domain, $10^{-6} \le q, p \le 10^{5}\,h\,\mathrm{Mpc}^{-1}$ (\texttt{FAST-PT}'s power-law extrapolation for $P_{22}$; the input-table range for $P_{13}$), so any residual difference is purely numerical.
We integrate with the deterministic, globally adaptive \texttt{Cuhre} algorithm~\cite{Berntsen:1991ciy,Hahn:2004fe} with a degree-13 cubature rule, in $(\ln q, \ln p)$ for $P_{22}$ and in $(\ln q, \hat{\mathbf{k}}\cdot\hat{\mathbf{q}})$ for $P_{13}$.

Three choices make the integration of an oscillating integrand tractable and accurate. First, $\Delta P_X$ is integrated directly in a cancellation-free form, e.g.
\begin{equation}
    P_L^\mathrm{nw}(k) P_L^\mathrm{nw}(q) \left[\delta P_\zeta(k) + \delta P_\zeta(q) + \delta P_\zeta(k)\,\delta P_\zeta(q)\right]
\end{equation}
for $P_{13}$, so that the error estimate refers to the feature part itself rather than to the difference of two large numbers. Second, the domain is split into regions not larger than half an oscillation period, graded towards the integrable singularity at $q = k$. Third, for the LIN template we write
\begin{equation}
    \sin\Phi(q)\sin\Phi(p) = \frac12\left\{\cos\tilde\omega_\mathrm{lin}(p-q) - \cos\left[\tilde\omega_\mathrm{lin}(p+q) + 2\phi_\mathrm{lin}\right]\right\} \,,
\end{equation}
with $\Phi(k) = \tilde\omega_\mathrm{lin} k + \phi_\mathrm{lin}$, and integrate each piece in the coordinates $(p+q,\, p-q)$, in which it oscillates along a single axis; at $\omega_\mathrm{lin} = 80$ this heavily reduces the number of integrand evaluations per $k$.

To make the tolerance meaningful for integrands that oscillate and change sign, each integral is normalised by a scale fixed in a cheap preliminary run, the $L^1$ norm of the no-wiggle integrand (multiplied by $A_X$ for the feature part), and \texttt{Cuhre} is run with a purely absolute tolerance of $10^{-8}$ ($P_{22}$) and $10^{-11}$ ($P_{13}$) in these units. Unlike a relative tolerance, this criterion stays well defined at the zeros of $\Delta P_X$. The \texttt{Cuhre} error estimates were checked against independent quadratures that share neither the integrator nor the coordinates (a composite Gauss--Legendre cubature of the unsplit $P_{22}$ integrand; an analytic angular integration followed by adaptive Gauss--Kronrod quadrature for $P_{13}$). All integrals converge at the 300 logarithmically spaced wavenumbers in $10^{-3} \le k \le 10\,h\,\mathrm{Mpc}^{-1}$, chosen to coincide with nodes of the \texttt{pbj} grid so that no interpolation enters the comparison. The direct integration costs a few CPU-hours per parameter set, against $\sim0.1\,$s for \texttt{FAST-PT}.

\subsection{Results and discussion}
\begin{figure}
    \centering
    \includegraphics[width=\linewidth]{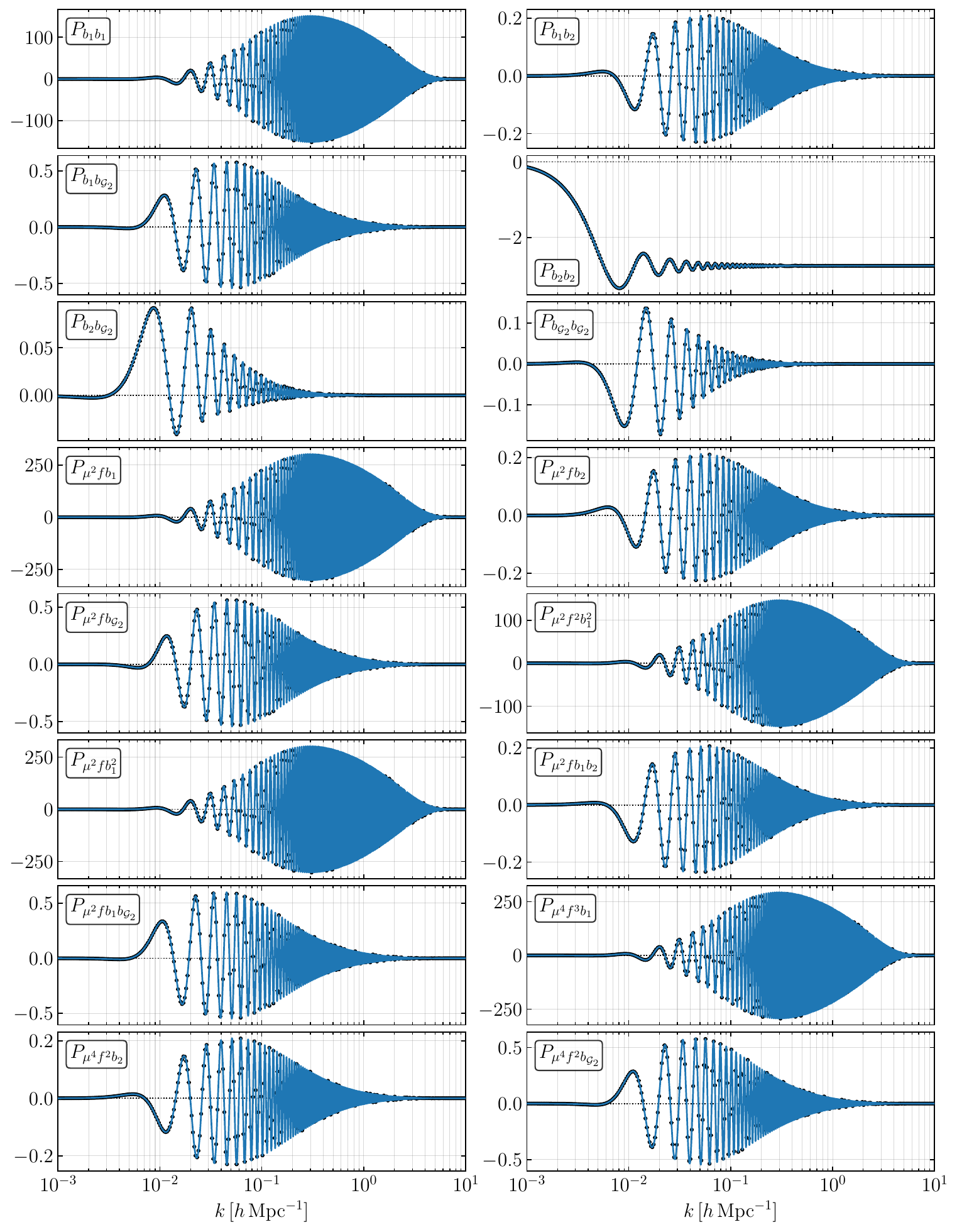}
    \caption{Feature part $\Delta P_X(k)$ of~\cref{eqn:loop_feature_part} for the first 16 independent $P_{22}$ integrals of~\cref{eqn:loop_decomp}, for the LIN template with $(A_\mathrm{lin}, \omega_\mathrm{lin}, \phi_\mathrm{lin}) = (0.05, 40, 0)$, in units of $(h^{-1}\,\mathrm{Mpc})^3$ at $z = 0$. Each panel is labelled by the coefficient $c_X$ multiplying the integral. Solid lines show \texttt{FAST-PT} with the default settings; black points show the direct two-dimensional integration with \texttt{Cuhre} at 300 wavenumbers in $10^{-3} \le k /(h\,\mathrm{Mpc}^{-1})\le 10$, whose $1\,\sigma$ integration errors are smaller than the points.}
    \label{fig:P1loop_LIN_1}
\end{figure}

\begin{figure}
    \centering
    \includegraphics[width=\linewidth]{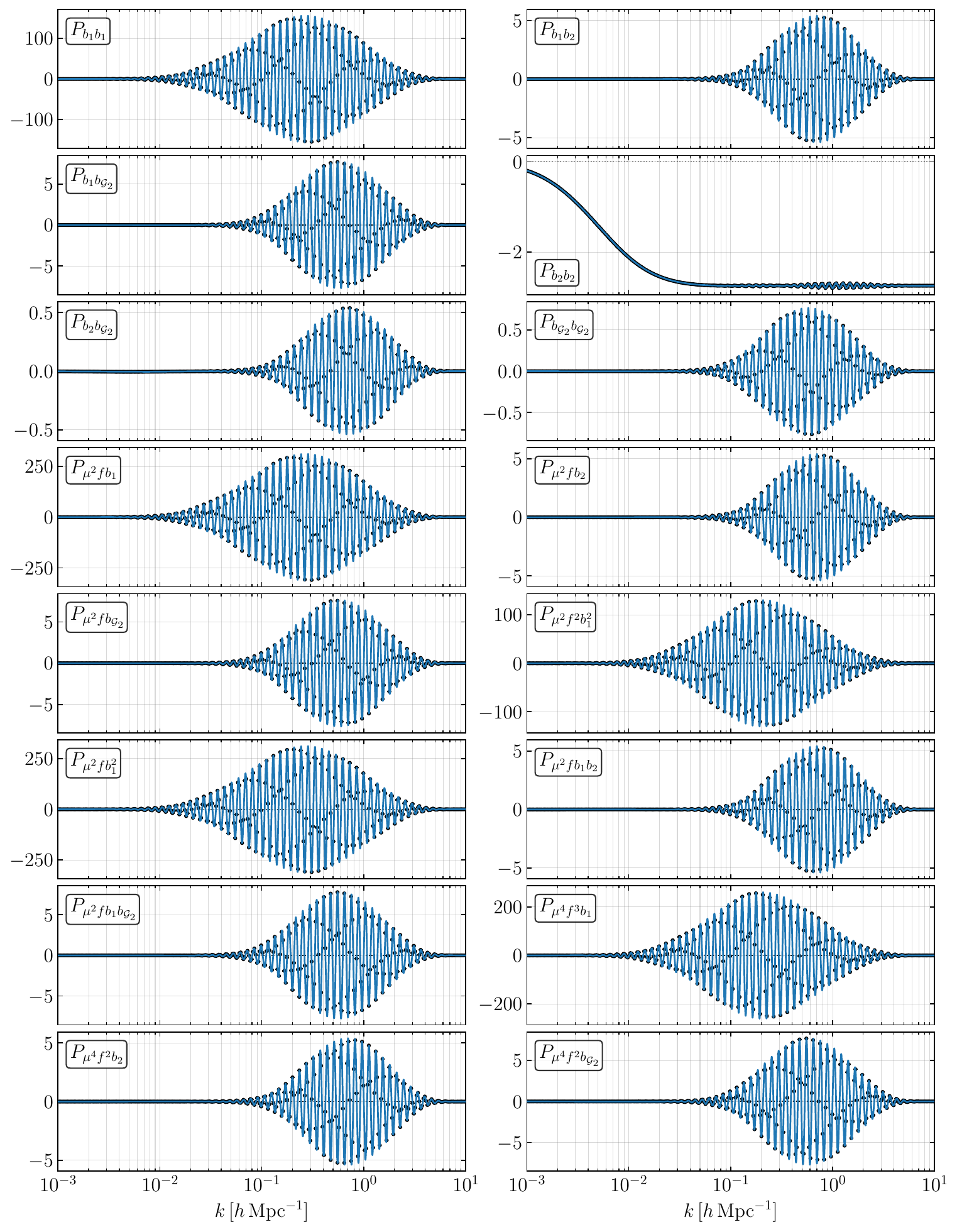}
    \caption{Same as~\cref{fig:P1loop_LIN_1} for the LOG template with $(A_\mathrm{log}, \omega_\mathrm{log}, \phi_\mathrm{log}) = (0.05, 40, 0)$.}
    \label{fig:P1loop_LOG_1}
\end{figure}

\begin{figure}
    \centering
    \includegraphics[width=\linewidth]{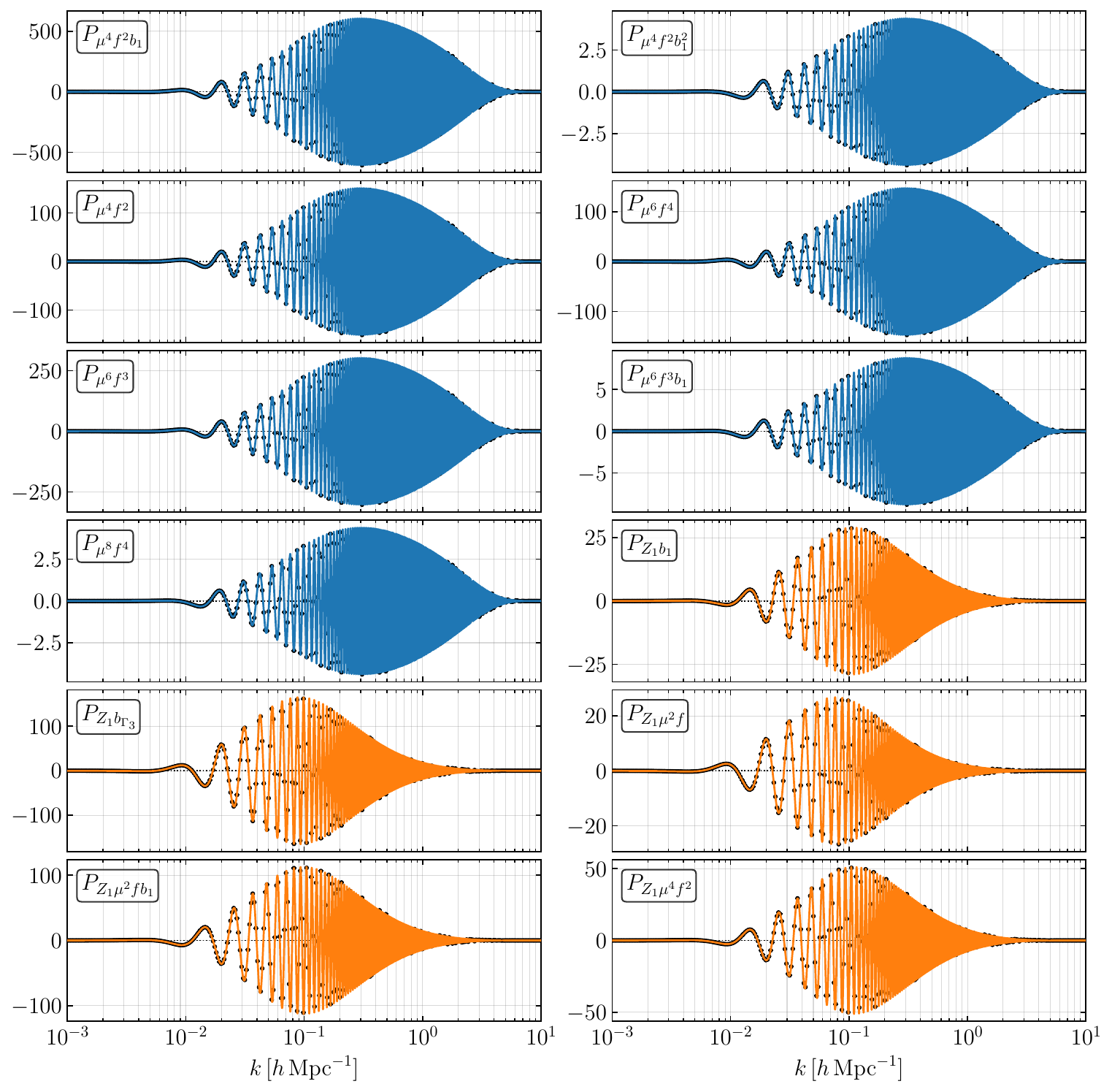}
    \caption{Same as~\cref{fig:P1loop_LIN_1} for the remaining seven independent $P_{22}$ integrals (blue) and the five independent $P_{13}$ integrals $\Delta P_{Z_1 Y}$ (orange).}
    \label{fig:P1loop_LIN_2}
\end{figure}

\begin{figure}
    \centering
    \includegraphics[width=\linewidth]{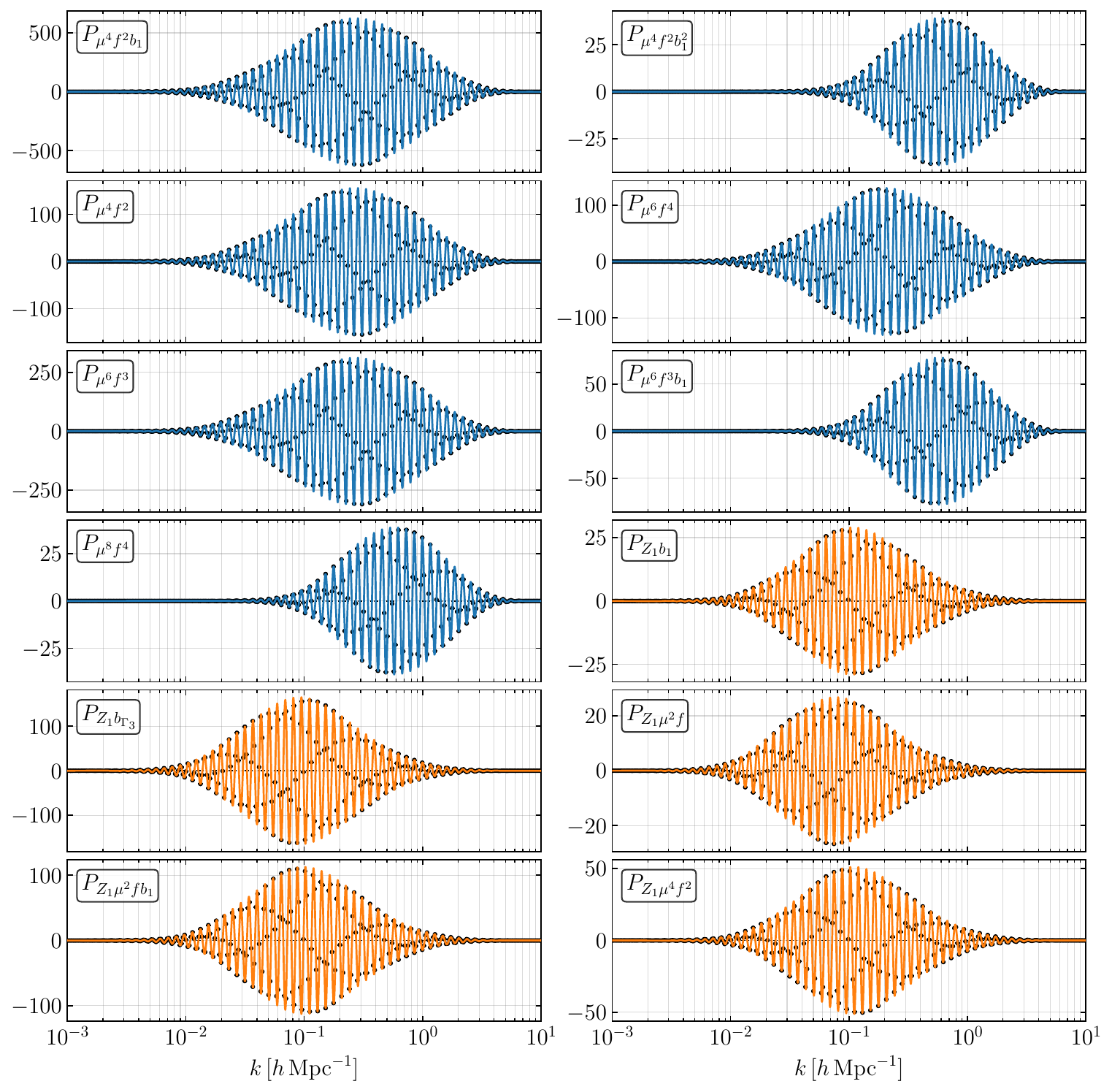}
    \caption{Same as~\cref{fig:P1loop_LIN_2} for the LOG template with $(A_\mathrm{log}, \omega_\mathrm{log}, \phi_\mathrm{log}) = (0.05, 40, 0)$.}
    \label{fig:P1loop_LOG_2}
\end{figure}

We adopt a flat $\Lambda$CDM cosmology with $h = 0.695$, $\omega_\mathrm{b} = 0.02224$, $\omega_\mathrm{c} = 0.1154$, $n_\mathrm{s} = 0.9632$, $A_\mathrm{s} = 2.2\times10^{-9}$ and massless neutrinos, and evaluate all integrals at $z = 0$, where nonlinear corrections, and hence any numerical difference, are largest. For the features we take $A_X = 0.05$ and $\phi_X = 0$, and test the LIN and LOG templates. These are representative of all four: BUMP and PSC share their oscillatory phases, modulated by a smooth envelope that does not make the comparison more demanding. We check all 28 independent integrals with the settings of \texttt{pbj} used in the analysis, and with a refined configuration, described below, that improves the agreement mostly at small scales, outside the range used in the analysis. \Cref{fig:P1loop_LIN_1,fig:P1loop_LIN_2,fig:P1loop_LOG_1,fig:P1loop_LOG_2} show the default case for an intermediate frequency, $\omega_X = 40$: the two methods are indistinguishable over the whole range of scales. \Cref{tab:loop_validation} quantifies the agreement for two extreme frequencies, $\omega_X = 10$ and $80$ ($\log_{10}\omega_X = 1$ and $\simeq 1.9$), which bracket almost the whole frequency range considered, normalising the residual of each integral to the linear feature signal $A_X P_L^\mathrm{nw}(k)$.

For the $P_{13}$ integrals, \texttt{pbj} reproduces the direct integration to $\lesssim 4\times10^{-7}$ for $0.0025 \le k \le 0.3\,h\,\mathrm{Mpc}^{-1}$ and to $\lesssim 2\times10^{-5}$ for all $k \le 1\,h\,\mathrm{Mpc}^{-1}$, in every case except LIN with $\omega_\mathrm{lin} = 80$, for which the input table under-samples the oscillation and the residual reaches $2\times10^{-3}$ and $2\times10^{-2}$, respectively (a four times finer table brings it back to $5\times10^{-7}$ for $k \le 0.3\,h\,\mathrm{Mpc}^{-1}$). The residual, which we trace to round-off in the FFT convolution and in the closed-form kernels at large $q/k$, grows to a few $\times10^{-2}$ ($3\times10^{-1}$ for LIN with $\omega_\mathrm{lin} = 80$) only at $k > 1\,h\,\mathrm{Mpc}^{-1}$, where the feature is exponentially damped.
For the $P_{22}$ integrals with the default settings (8\,000 table nodes with $\Delta\ln k = 1.8\times10^{-3}$, zero padding over $1.8$ e-folds, and a Fourier filter acting on the outer $75\%$ of the modes), the residual for $0.0025 \le k \le 0.3\,h\,\mathrm{Mpc}^{-1}$ is $\lesssim 5\times10^{-8}$ for $\omega_X = 10$ and grows with frequency up to $\sim 2\times10^{-3}$ for $\omega_\mathrm{lin} = 80$. Two effects are responsible: the short zero padding, which dominates at low $k$ and for LOG, and, for LIN, the sampling of the oscillation. The input table has fewer than two nodes per period above $k_\mathrm{Nyq} = \pi/(\tilde\omega_\mathrm{lin}\Delta\ln k)$, i.e.\ $\simeq 1.6\,h\,\mathrm{Mpc}^{-1}$ for $\omega_\mathrm{lin} = 80$, and the Fourier filter starts suppressing the feature already at $\simeq k_\mathrm{Nyq}/4$; as a consequence, above $k \simeq 0.4$--$0.7\,h\,\mathrm{Mpc}^{-1}$ the default configuration does not reproduce the LIN feature for $\omega_\mathrm{lin} \ge 50$, and above $k \simeq 3\,h\,\mathrm{Mpc}^{-1}$ even for $\omega_\mathrm{lin} = 10$. In the refined configuration, the input table is finer (by factors 4 and 32 for $\omega_\mathrm{lin} = 10,\,80$) and the padding extends over 29 e-folds; this reduces the residual to $\lesssim 5\times10^{-9}$ for $k \le 0.3\,h\,\mathrm{Mpc}^{-1}$ and $\lesssim 2\times10^{-7}$ for $k \le 1\,h\,\mathrm{Mpc}^{-1}$ in all cases, within the precision of the reference.

Over the range of the data, $0.0025 \le k \le 0.3\,h\,\mathrm{Mpc}^{-1}$, the default configuration therefore reproduces every one-loop integral to better than $3\times10^{-3}$ of the linear feature signal (worst case LIN with $\omega_\mathrm{lin} = 80$, and $\lesssim 10^{-6}$ in all other cases). Since the coefficients $c_X$ and $d_Y$ are of order unity, for $A_X = 0.05$ this is at most ${\sim}10^{-4}$ of the galaxy power spectrum, several orders of magnitude below the statistical uncertainties of \textit{BOSS} and \textit{eBOSS}. The default configuration is thus adequate for the present analysis, while the refined one, which matters only at smaller scales, would considerably lengthen the runs.

\begin{table}[t]
    \centering
    \small
    \begin{tabular}{l c c c c c c c}
    \noalign{\vskip 3pt}\hline\noalign{\vskip 1.5pt}\hline\noalign{\vskip 5pt}
     & & \multicolumn{3}{c}{\boldmath$0.0025 \le k \le 0.3$} & \multicolumn{3}{c}{\boldmath$0.3 < k \le 1$} \\
    {\bf Template} & $\omega_X$ & $P_{Z_1 Y}$ (\texttt{pbj}) & $P_X$ (\texttt{pbj}) & $P_X$ (refined) & $P_{Z_1 Y}$ (\texttt{pbj}) & $P_X$ (\texttt{pbj}) & $P_X$ (refined) \\
    \noalign{\vskip 3pt}\hline\noalign{\vskip 3pt}
    LIN & 10 & $4\times10^{-7}$ & $5\times10^{-8}$ & $6\times10^{-10}$ & $1\times10^{-5}$ & $6\times10^{-7}$ & $6\times10^{-9}$ \\
    LIN & 80 & $2\times10^{-3}$ & $2\times10^{-3}$ & $6\times10^{-10}$ & $2\times10^{-2}$ & $4\times10^{1}$ & $2\times10^{-9}$ \\
    LOG & 10 & $4\times10^{-7}$ & $5\times10^{-9}$ & $3\times10^{-9}$ & $2\times10^{-5}$ & $2\times10^{-7}$ & $2\times10^{-7}$ \\
    LOG & 80 & $3\times10^{-7}$ & $2\times10^{-7}$ & $1\times10^{-9}$ & $2\times10^{-5}$ & $2\times10^{-7}$ & $7\times10^{-8}$ \\
    \hline\noalign{\vskip 1.5pt}\hline
    \end{tabular}
    \caption{Maximum deviation of \texttt{FAST-PT} from the direct integration of the feature part of the one-loop integrals, $\max|\Delta P^{\texttt{FAST-PT}} - \Delta P^{\texttt{Cuhre}}|/(A_X P_L^\mathrm{nw})$ over all integrals of each type, for $A_X = 0.05$ and $\phi_X = 0$ at $z = 0$, with $k$ in $h\,\mathrm{Mpc}^{-1}$. Columns 3--5 cover the range of the data; the maxima are unchanged if the range is extended down to $k = 10^{-3}\,h\,\mathrm{Mpc}^{-1}$. ``\texttt{pbj}'' denotes the default settings used in the analysis, ``refined'' the extended padding and, for LIN, the finer input table described in the text.}
    \label{tab:loop_validation}
\end{table}

\FloatBarrier
\bibliographystyle{JHEP}
\bibliography{bibliography}

\end{document}